\documentclass[11pt]{article}

\usepackage{fullpage}
\usepackage{amsfonts}

\usepackage{alltt}
\def\R{\mathbb{R}}
\newsavebox{\savepar}

\ifx\undefined\psfig\else \fi

\edef\psfigRestoreAt{\catcode`@=\number\catcode`@\relax}
\catcode`\@=11\relax
\newwrite\@unused
\def\typeout#1{{\let\protect\string\immediate\write\@unused{#1}}}
\def\figurepath{./}

\def\@nnil{\@nil}
\def\@empty{}
\def\@psdonoop#1\@@#2#3{}
\def\@psdo#1:=#2\do#3{\edef\@psdotmp{#2}\ifx\@psdotmp\@empty \else
    \expandafter\@psdoloop#2,\@nil,\@nil\@@#1{#3}\fi}
\def\@psdoloop#1,#2,#3\@@#4#5{\def#4{#1}\ifx #4\@nnil \else
       #5\def#4{#2}\ifx #4\@nnil \else#5\@ipsdoloop #3\@@#4{#5}\fi\fi}
\def\@ipsdoloop#1,#2\@@#3#4{\def#3{#1}\ifx #3\@nnil 
       \let\@nextwhile=\@psdonoop \else
      #4\relax\let\@nextwhile=\@ipsdoloop\fi\@nextwhile#2\@@#3{#4}}
\def\@tpsdo#1:=#2\do#3{\xdef\@psdotmp{#2}\ifx\@psdotmp\@empty \else
    \@tpsdoloop#2\@nil\@nil\@@#1{#3}\fi}
\def\@tpsdoloop#1#2\@@#3#4{\def#3{#1}\ifx #3\@nnil 
       \let\@nextwhile=\@psdonoop \else
      #4\relax\let\@nextwhile=\@tpsdoloop\fi\@nextwhile#2\@@#3{#4}}
\newread\ps@stream
\newif\ifnot@eof       
\newif\if@noisy        
\newif\if@atend        
\newif\if@psfile       
{\catcode`\%=12\global\gdef\epsf@start{
\def\epsf@PS{PS}
\def\epsf@getbb#1{%
\openin\ps@stream=#1
\ifeof\ps@stream\typeout{Error, File #1 not found}\else
   {\not@eoftrue \chardef\other=12
    \def\do##1{\catcode`##1=\other}\dospecials \catcode`\ =10
    \loop
       \if@psfile
	  \read\ps@stream to \epsf@fileline
       \else{
	  \obeyspaces
          \read\ps@stream to \epsf@tmp\global\let\epsf@fileline\epsf@tmp}
       \fi
       \ifeof\ps@stream\not@eoffalse\else
       \if@psfile\else
       \expandafter\epsf@test\epsf@fileline:. \\%
       \fi
          \expandafter\epsf@aux\epsf@fileline:. \\%
       \fi
   \ifnot@eof\repeat
   }\closein\ps@stream\fi}%
\long\def\epsf@test#1#2#3:#4\\{\def\epsf@testit{#1#2}
			\ifx\epsf@testit\epsf@start\else
\typeout{Warning! File does not start with `\epsf@start'.  It may not be a PostScript file.}
			\fi
			\@psfiletrue} 
{\catcode`\%=12\global\let\epsf@percent=
\long\def\epsf@aux#1#2:#3\\{\ifx#1\epsf@percent
   \def\epsf@testit{#2}\ifx\epsf@testit\epsf@bblit
	\@atendfalse
        \epsf@atend #3 . \\%
	\if@atend	
	   \if@verbose{
		\typeout{psfig: found `(atend)'; continuing search}
	   }\fi
        \else
        \epsf@grab #3 . . . \\%
        \not@eoffalse
        \global\no@bbfalse
        \fi
   \fi\fi}%
\def\epsf@grab #1 #2 #3 #4 #5\\{%
   \global\def\epsf@llx{#1}\ifx\epsf@llx\empty
      \epsf@grab #2 #3 #4 #5 .\\\else
   \global\def\epsf@lly{#2}%
   \global\def\epsf@urx{#3}\global\def\epsf@ury{#4}\fi}%
\def\epsf@atendlit{(atend)} 
\def\epsf@atend #1 #2 #3\\{%
   \def\epsf@tmp{#1}\ifx\epsf@tmp\empty
      \epsf@atend #2 #3 .\\\else
   \ifx\epsf@tmp\epsf@atendlit\@atendtrue\fi\fi}

\chardef\letter = 11
\chardef\other = 12

\newif \ifdebug 
\newif\ifc@mpute 
\c@mputetrue 

\let\then = \relax
\def\r@dian{pt }
\let\r@dians = \r@dian
\let\dimensionless@nit = \r@dian
\let\dimensionless@nits = \dimensionless@nit
\def\internal@nit{sp }
\let\internal@nits = \internal@nit
\newif\ifstillc@nverging
\def \Mess@ge #1{\ifdebug \then \message {#1} \fi}

{ 
	\catcode `\@ = \letter
	\gdef \nodimen {\expandafter \n@dimen \the \dimen}
	\gdef \term #1 #2 #3%
	       {\edef \t@ {\the #1}
		\edef \t@@ {\expandafter \n@dimen \the #2\r@dian}%
		\t@rm {\t@} {\t@@} {#3}%
	       }
	\gdef \t@rm #1 #2 #3%
	       {{%
		\count 0 = 0
		\dimen 0 = 1 \dimensionless@nit
		\dimen 2 = #2\relax
		\Mess@ge {Calculating term #1 of \nodimen 2}%
		\loop
		\ifnum	\count 0 < #1
		\then	\advance \count 0 by 1
			\Mess@ge {Iteration \the \count 0 \space}%
			\Multiply \dimen 0 by {\dimen 2}%
			\Mess@ge {After multiplication, term = \nodimen 0}%
			\Divide \dimen 0 by {\count 0}%
			\Mess@ge {After division, term = \nodimen 0}%
		\repeat
		\Mess@ge {Final value for term #1 of 
				\nodimen 2 \space is \nodimen 0}%
		\xdef \Term {#3 = \nodimen 0 \r@dians}%
		\aftergroup \Term
	       }}
	\catcode `\p = \other
	\catcode `\t = \other
	\gdef \n@dimen #1pt{#1} 
}

\def \Divide #1by #2{\divide #1 by #2} 

\def \Multiply #1by #2
       {{
	\count 0 = #1\relax
	\count 2 = #2\relax
	\count 4 = 65536
	\Mess@ge {Before scaling, count 0 = \the \count 0 \space and
			count 2 = \the \count 2}%
	\ifnum	\count 0 > 32767 
	\then	\divide \count 0 by 4
		\divide \count 4 by 4
	\else	\ifnum	\count 0 < -32767
		\then	\divide \count 0 by 4
			\divide \count 4 by 4
		\else
		\fi
	\fi
	\ifnum	\count 2 > 32767 
	\then	\divide \count 2 by 4
		\divide \count 4 by 4
	\else	\ifnum	\count 2 < -32767
		\then	\divide \count 2 by 4
			\divide \count 4 by 4
		\else
		\fi
	\fi
	\multiply \count 0 by \count 2
	\divide \count 0 by \count 4
	\xdef \product {#1 = \the \count 0 \internal@nits}%
	\aftergroup \product
       }}

\def\r@duce{\ifdim\dimen0 > 90\r@dian \then   
		\multiply\dimen0 by -1
		\advance\dimen0 by 180\r@dian
		\r@duce
	    \else \ifdim\dimen0 < -90\r@dian \then  
		\advance\dimen0 by 360\r@dian
		\r@duce
		\fi
	    \fi}

\def\Sine#1%
       {{%
	\dimen 0 = #1 \r@dian
	\r@duce
	\ifdim\dimen0 = -90\r@dian \then
	   \dimen4 = -1\r@dian
	   \c@mputefalse
	\fi
	\ifdim\dimen0 = 90\r@dian \then
	   \dimen4 = 1\r@dian
	   \c@mputefalse
	\fi
	\ifdim\dimen0 = 0\r@dian \then
	   \dimen4 = 0\r@dian
	   \c@mputefalse
	\fi
	\ifc@mpute \then
		\divide\dimen0 by 180
		\dimen0=3.141592654\dimen0
		\dimen 2 = 3.1415926535897963\r@dian 
		\divide\dimen 2 by 2 
		\Mess@ge {Sin: calculating Sin of \nodimen 0}%
		\count 0 = 1 
		\dimen 2 = 1 \r@dian 
		\dimen 4 = 0 \r@dian 
		\loop
			\ifnum	\dimen 2 = 0 
			\then	\stillc@nvergingfalse 
			\else	\stillc@nvergingtrue
			\fi
			\ifstillc@nverging 
			\then	\term {\count 0} {\dimen 0} {\dimen 2}%
				\advance \count 0 by 2
				\count 2 = \count 0
				\divide \count 2 by 2
				\ifodd	\count 2 
				\then	\advance \dimen 4 by \dimen 2
				\else	\advance \dimen 4 by -\dimen 2
				\fi
		\repeat
	\fi		
			\xdef \sine {\nodimen 4}%
       }}

\def\Cosine#1{\ifx\sine\UnDefined\edef\Savesine{\relax}\else
		             \edef\Savesine{\sine}\fi
	{\dimen0=#1\r@dian\multiply\dimen0 by -1
	 \advance\dimen0 by 90\r@dian
	 \Sine{\nodimen 0}
	 \xdef\cosine{\sine}
	 \xdef\sine{\Savesine}}}	      

\def\psdraft{
	\def\@psdraft{0}
}
\def\psfull{
	\def\@psdraft{100}
}

\psfull

\newif\if@draftbox
\def\psnodraftbox{
	\@draftboxfalse
}
\@draftboxtrue

\newif\if@prologfile
\newif\if@postlogfile
\def\pssilent{
	\@noisyfalse
}
\def\psnoisy{
	\@noisytrue
}
\psnoisy
\newif\if@bbllx
\newif\if@bblly
\newif\if@bburx
\newif\if@bbury
\newif\if@height
\newif\if@width
\newif\if@rheight
\newif\if@rwidth
\newif\if@angle
\newif\if@clip
\newif\if@verbose
\newif\if@scale
\def\@p@@sclip#1{\@cliptrue}

\def\@p@@sfile#1{\def\@p@sfile{null}%
	        \openin1=#1
		\ifeof1\closein1%
		       \openin1=\figurepath#1
			\ifeof1\typeout{Error, File #1 not found}
			   \if@bbllx\if@bblly\if@bburx\if@bbury
			      \def\@p@sfile{#1}%
			   \fi\fi\fi\fi
			\else\closein1
			    \edef\@p@sfile{\figurepath#1}%
                        \fi%
		 \else\closein1%
		       \def\@p@sfile{#1}%
		 \fi}
\def\@p@@sfigure#1{\def\@p@sfile{null}%
	        \openin1=#1
		\ifeof1\closein1%
		       \openin1=\figurepath#1
			\ifeof1\typeout{Error, File #1 not found}
			   \if@bbllx\if@bblly\if@bburx\if@bbury
			      \def\@p@sfile{#1}%
			   \fi\fi\fi\fi
			\else\closein1
			    \def\@p@sfile{\figurepath#1}%
                        \fi%
		 \else\closein1%
		       \def\@p@sfile{#1}%
		 \fi}

\def\@p@@sbbllx#1{
		\@bbllxtrue
		\dimen100=#1
		\edef\@p@sbbllx{\number\dimen100}
}
\def\@p@@sbblly#1{
		\@bbllytrue
		\dimen100=#1
		\edef\@p@sbblly{\number\dimen100}
}
\def\@p@@sbburx#1{
		\@bburxtrue
		\dimen100=#1
		\edef\@p@sbburx{\number\dimen100}
}
\def\@p@@sbbury#1{
		\@bburytrue
		\dimen100=#1
		\edef\@p@sbbury{\number\dimen100}
}
\def\@p@@sheight#1{
		\@heighttrue
		\dimen100=#1
   		\edef\@p@sheight{\number\dimen100}
}
\def\@p@@swidth#1{
		\@widthtrue
		\dimen100=#1
		\edef\@p@swidth{\number\dimen100}
}
\def\@p@@srheight#1{
		\@rheighttrue
		\dimen100=#1
		\edef\@p@srheight{\number\dimen100}
}
\def\@p@@srwidth#1{
		\@rwidthtrue
		\dimen100=#1
		\edef\@p@srwidth{\number\dimen100}
}
\def\@p@@sangle#1{
		\@angletrue
		\edef\@p@sangle{#1} 
}
\def\@p@@ssilent#1{ 
		\@verbosefalse
}
\def\@p@@sscale#1{
		\def\@p@scale{#1}
		\@scaletrue
}
\def\@p@@sprolog#1{\@prologfiletrue\def\@prologfileval{#1}}
\def\@p@@spostlog#1{\@postlogfiletrue\def\@postlogfileval{#1}}
\def\@cs@name#1{\csname #1\endcsname}
\def\@setparms#1=#2,{\@cs@name{@p@@s#1}{#2}}
\def\ps@init@parms{
		\@bbllxfalse \@bbllyfalse
		\@bburxfalse \@bburyfalse
		\@heightfalse \@widthfalse
		\@rheightfalse \@rwidthfalse
		\@scalefalse
		\def\@p@sbbllx{}\def\@p@sbblly{}
		\def\@p@sbburx{}\def\@p@sbbury{}
		\def\@p@sheight{}\def\@p@swidth{}
		\def\@p@srheight{}\def\@p@srwidth{}
		\def\@p@sangle{0}
		\def\@p@sfile{}
		\def\@p@scost{10}
		\def\@sc{}
		\@prologfilefalse
		\@postlogfilefalse
		\@clipfalse
		\if@noisy
			\@verbosetrue
		\else
			\@verbosefalse
		\fi
}
\def\parse@ps@parms#1{
	 	\@psdo\@psfiga:=#1\do
		   {\expandafter\@setparms\@psfiga,}}
\newif\ifno@bb
\def\bb@missing{
	\if@verbose{
		\typeout{psfig: searching \@p@sfile \space  for bounding box}
	}\fi
	\no@bbtrue
	\epsf@getbb{\@p@sfile}
        \ifno@bb \else \bb@cull\epsf@llx\epsf@lly\epsf@urx\epsf@ury\fi
}	
\def\bb@cull#1#2#3#4{
	\dimen100=#1 bp\edef\@p@sbbllx{\number\dimen100}
	\dimen100=#2 bp\edef\@p@sbblly{\number\dimen100}
	\dimen100=#3 bp\edef\@p@sbburx{\number\dimen100}
	\dimen100=#4 bp\edef\@p@sbbury{\number\dimen100}
	\no@bbfalse
}

\newdimen\p@intvaluex
\newdimen\p@intvaluey
\newdimen\@ffsetvalue
\newdimen\x@ffsetvalue
\newdimen\y@ffsetvalue

\def\compute@offset#1#2{{\dimen0=#1 sp\dimen1=#2 sp
			\advance\dimen1 by -\dimen0
			\dimen1=\sine\dimen1
			\dimen0=\cosine\dimen1
			\ifdim\dimen0<0sp \dimen1=0sp \fi
			\global\@ffsetvalue=\dimen1}}

\def\rotate@#1#2{{\dimen0=#1 sp\dimen1=#2 sp
		  \global\p@intvaluex=\cosine\dimen0
		  \dimen3=\sine\dimen1
		  \global\advance\p@intvaluex by -\dimen3
		  \global\p@intvaluey=\sine\dimen0
		  \dimen3=\cosine\dimen1
		  \global\advance\p@intvaluey by \dimen3
		  }}
\def\compute@bb{
		\no@bbfalse
		\if@bbllx \else \no@bbtrue \fi
		\if@bblly \else \no@bbtrue \fi
		\if@bburx \else \no@bbtrue \fi
		\if@bbury \else \no@bbtrue \fi
		\ifno@bb \bb@missing \fi
		\ifno@bb \typeout{FATAL ERROR: no bb supplied or found}
			\no-bb-error
		\fi
		\if@angle 
			\Sine{\@p@sangle}\Cosine{\@p@sangle}
			\compute@offset{\@p@sbblly}{\@p@sbbury}
			\x@ffsetvalue=\@ffsetvalue
			\compute@offset{\@p@sbburx}{\@p@sbbllx}
			\y@ffsetvalue=\@ffsetvalue

			\rotate@{\@p@sbbllx}{\@p@sbblly}
			\advance\p@intvaluex by -\x@ffsetvalue
			\advance\p@intvaluey by -\y@ffsetvalue
			\edef\@p@sbbllx{\number\p@intvaluex}
			\edef\@p@sbblly{\number\p@intvaluey}

			\rotate@{\@p@sbburx}{\@p@sbbury}
			\advance\p@intvaluex by \x@ffsetvalue
			\advance\p@intvaluey by \y@ffsetvalue
			\edef\@p@sbburx{\number\p@intvaluex}
			\edef\@p@sbbury{\number\p@intvaluey}
			{
			 \count0=\@p@sbbllx \count1=\@p@sbblly
		 	 \count2=\@p@sbburx \count3=\@p@sbbury
			 \dimen0=\@p@sbbllx sp\dimen1=\@p@sbblly sp
		 	 \dimen2=\@p@sbburx sp\dimen3=\@p@sbbury sp
			 \dimen203=\dimen2 \advance\dimen203 by -\dimen0
			 \dimen204=\dimen3 \advance\dimen204 by -\dimen1
			 \ifdim\dimen203<0sp 
			      \count203=\count2 \count2=\count0 
			      \count0=\count203 
			      \global\edef\@p@sbbllx{\number\count0}
			      \global\edef\@p@sbburx{\number\count2}
			 \fi
			 \ifdim\dimen204<0sp 
			       \count204=\count3
			       \count3=\count1
			       \count1=\count204
			       \global\edef\@p@sbblly{\number\count1}
			       \global\edef\@p@sbbury{\number\count3}
			 \fi
			}
		\fi
		\count203=\@p@sbburx
		\count204=\@p@sbbury
		\advance\count203 by -\@p@sbbllx
		\advance\count204 by -\@p@sbblly
		\edef\@bbw{\number\count203}
		\edef\@bbh{\number\count204}
}
\def\in@hundreds#1#2#3{\count240=#2 \count241=#3
		     \count100=\count240	
		     \divide\count100 by \count241
		     \count101=\count100
		     \multiply\count101 by \count241
		     \advance\count240 by -\count101
		     \multiply\count240 by 10
		     \count101=\count240	
		     \divide\count101 by \count241
		     \count102=\count101
		     \multiply\count102 by \count241
		     \advance\count240 by -\count102
		     \multiply\count240 by 10
		     \count102=\count240	
		     \divide\count102 by \count241
		     \count200=#1\count205=0
		     \count201=\count200
			\multiply\count201 by \count100
		 	\advance\count205 by \count201
		     \count201=\count200
			\divide\count201 by 10
			\multiply\count201 by \count101
			\advance\count205 by \count201
		     \count201=\count200
			\divide\count201 by 100
			\multiply\count201 by \count102
			\advance\count205 by \count201
		     \edef\@result{\number\count205}
}
\def\@ScaleInHundreds#1{
		\in@hundreds{#1}{\@p@scale}{100}
		\edef#1{\@result}
}
\def\compute@wfromh{
		\in@hundreds{\@p@sheight}{\@bbw}{\@bbh}
		\edef\@p@swidth{\@result}
}
\def\compute@hfromw{
		\in@hundreds{\@p@swidth}{\@bbh}{\@bbw}
		\edef\@p@sheight{\@result}
}
\def\compute@handw{
		\if@height 
			\if@width
			\else
				\compute@wfromh
			\fi
		\else 
			\if@width
				\compute@hfromw
			\else
				\edef\@p@sheight{\@bbh}
				\edef\@p@swidth{\@bbw}
			\fi
		\fi
}
\def\compute@resv{
		\if@rheight \else \edef\@p@srheight{\@p@sheight} \fi
		\if@rwidth \else \edef\@p@srwidth{\@p@swidth} \fi
}
\def\compute@sizes{
	\compute@bb
	\compute@handw
	\compute@resv
}
\def\psfig#1{\vbox {
	\ps@init@parms
	\parse@ps@parms{#1}
	\compute@sizes
	\if@scale
                \if@verbose
                        \typeout{psfig: scaling by \@p@scale}
                \fi
                \@ScaleInHundreds{\@p@swidth}
                \@ScaleInHundreds{\@p@sheight}
                \@ScaleInHundreds{\@p@srwidth}
                \@ScaleInHundreds{\@p@srheight}
        \fi
	\ifnum\@p@scost<\@psdraft{
		\if@verbose{
			\typeout{psfig: including \@p@sfile \space }
		}\fi
		\special{ps::[begin] 	\@p@swidth \space \@p@sheight \space
				\@p@sbbllx \space \@p@sbblly \space
				\@p@sbburx \space \@p@sbbury \space
				startTexFig \space }
		\if@angle
			\special {ps:: \@p@sangle \space rotate \space} 
		\fi
		\if@clip{
			\if@verbose{
				\typeout{(clip)}
			}\fi
			\special{ps:: doclip \space }
		}\fi
		\if@prologfile
		    \special{ps: plotfile \@prologfileval \space } \fi
		\special{ps: plotfile \@p@sfile \space }
		\if@postlogfile
		    \special{ps: plotfile \@postlogfileval \space } \fi
		\special{ps::[end] endTexFig \space }
		\vbox to \@p@srheight true sp{
			\hbox to \@p@srwidth true sp{
				\hss
			}
		\vss
		}
	}\else{
		\if@draftbox{		
			\hbox{\fbox{\vbox to \@p@srheight true sp{
			\vss
			\hbox to \@p@srwidth true sp{ \hss \@p@sfile \hss }
			\vss
			}}}
		}\else{
			\vbox to \@p@srheight true sp{
			\vss
			\hbox to \@p@srwidth true sp{\hss}
			\vss
			}
		}\fi

	}\fi
}}
\def\psglobal{\typeout{psfig: PSGLOBAL is OBSOLETE; use psprint -m instead}}
\psfigRestoreAt

\newif\ifpdf
\ifx\pdfoutput\undefined
  \pdffalse
\else
  \pdfoutput=1
  \pdftrue
\fi

\ifpdf
  \usepackage[pdftex]{graphicx}
  \usepackage[pdftex]{color}
  \DeclareGraphicsExtensions{.pdf,.png,.jpg}
\else
  \usepackage[dvips]{graphicx}
  \usepackage[dvips]{color}
  \DeclareGraphicsExtensions{.eps,.epsi,.ps}
\fi

\usepackage{times}

\def\midv{\mathop{\,|\,}}

\long\def\cbk#1{{\color{red}[CBK: #1]}}
\newlength\colwidth \setlength\colwidth{3.25in}

\title{Sampling Strategies for Mining in Data-Scarce Domains}

\author{Naren Ramakrishnan \\
Department of Computer Science\\
Virginia Tech, VA 24061\\
Tel: (540) 231-8451\\
Email: naren@cs.vt.edu
\and
Chris Bailey-Kellogg\\
Department of Computer Sciences\\
Purdue University, IN 47907\\
Tel: (765) 494-9025\\
Email: cbk@cs.purdue.edu}

\date{}
\begin{document}

\maketitle
\begin{abstract}
\noindent
Data mining has traditionally focused on the task of drawing
inferences from large datasets.  However, many scientific and
engineering domains, such as fluid dynamics and aircraft design, are
characterized by {\em scarce} data, due to the expense and
complexity of associated experiments and simulations.  In such
data-scarce domains, it is advantageous to focus the data collection
effort on only those regions deemed most important to support a
particular data mining objective.  This paper describes a mechanism
that interleaves bottom-up data mining, to uncover multi-level
structures in spatial data, with top-down sampling, to clarify
difficult decisions in the mining process.  The mechanism exploits
relevant physical properties, such as continuity, correspondence, and
locality, in a unified framework. This leads to effective mining and
sampling decisions that are explainable in terms of domain knowledge
and data characteristics.  This approach is demonstrated in two
diverse applications --- mining pockets in spatial data, and
qualitative determination of Jordan forms of matrices.
\end{abstract}

\section{Introduction}
\begin{figure}
\begin{center}
\includegraphics[width=4.5in]{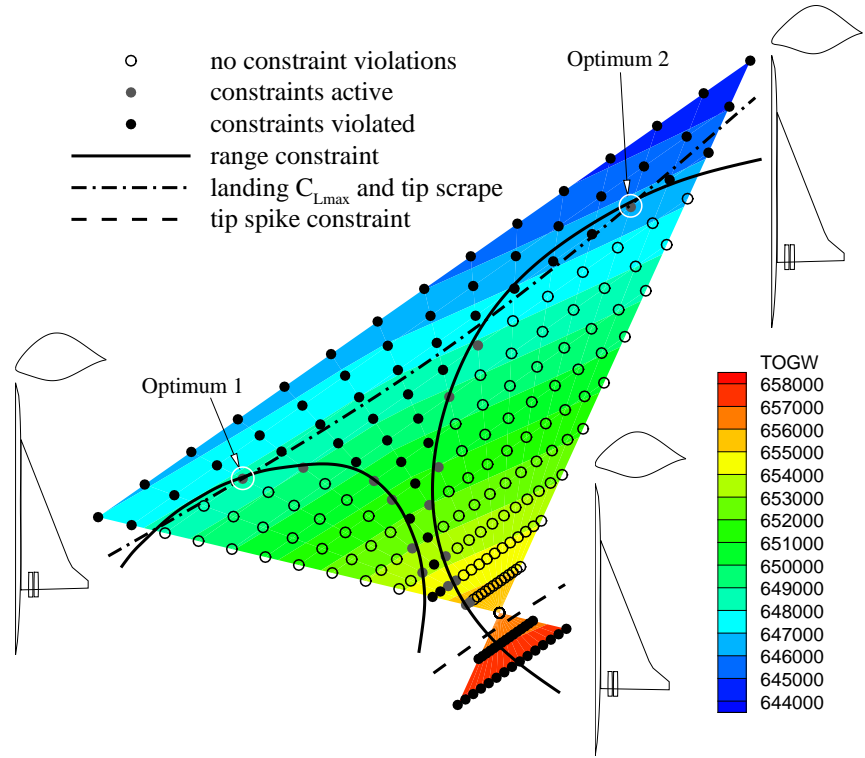}
\end{center}
\vspace*{-\baselineskip}
\caption{A pocket in an aircraft design space viewed as a slice
through three design points~\cite{vizcraft} (courtesy Layne T. Watson).}
\label{fig:aircraft}
\end{figure}

A number of important scientific and engineering applications, such as
fluid dynamics simulation and aircraft design, require analysis of 
spatially-distributed data from expensive experiments
and/or complex simulations demanding days, weeks, or even years on
petaflops-class computing systems.  For example, 
consider the conceptual design of a high-speed civil transport (HSCT), 
which involves the disciplines of aerodynamics, structures,
controls (mission-related), and propulsion. 80\% of the 
aircraft lifecycle cost is determined at this stage.
Fig.~\ref{fig:aircraft} shows a cross-section of the design space for
such a problem involving 29 design
variables with 68 constraints~\cite{vizcraft}. 
Frequently, the engineer will change some aspect of
a nominal design point, and run a simulation to see how the change
affects the objective function and various constraints dealing with
aircraft geometry and performance/aerodynamics. Or the design
process is made configurable, so the engineer can concentrate
on accurately modeling some aspect (e.g., the interaction between the
wing root and the fuselage) while replacing the remainder of the design
with fixed boundary conditions surrounding the focal area. Both these
approaches are inadequate for exploring such large high-dimensional design spaces,
even at low fidelity.  Ideally, the design engineer would like a
high-level mining system to identify the {\it pockets} that contain
good designs and which merit further consideration; traditional tools
from optimization and approximation theory can then be applied to
fine-tune such preliminary analyses. 

Three important characteristics distinguish such applications.
First, they are characterized not by an abundance of data, but
rather by a scarcity of data (owing to the cost and time involved in
conducting simulations). Second, the
computational scientist has complete control over the data acquisition
process (e.g.\ regions of the design space where data can be
collected), especially via computer simulations. 
And finally, there exists significant domain knowledge
in the form of physical properties such as continuity, correspondence, and
locality. It is natural therefore to use such information to focus data
collection for data mining. In this paper, we are interested in 
the question: `Given a simulation code, knowledge of physical properties, and a data
mining goal, at what points should data be collected?'

By suitably formulating an objective function and constraints around this question, we can
pose it as a problem of minimizing the number of samples needed for data mining.
Such a combination of \{data-scarcity + control over data collection + 
need to exploit domain knowledge\} characterizes many important 
computational science applications.
Data mining is now recognized as a key solution approach
for such applications, supporting analysis, visualization, 
and design tasks~\cite{naren-ayg-advances}. It serves a primary 
role in many domains (e.g., microarray bioinformatics) and a complementary role in 
others, by augmenting traditional techniques from numerical analysis, 
statistics, and machine learning. 

The goal of this paper is to describe focused sampling strategies  
for mining scientific data. Our approach is based on the spatial aggregation 
language (SAL)~\cite{bailey-kellogg96}, which supports construction 
of data interpretation and control design applications for 
spatially-distributed physical systems. Used as a basis for describing data mining
algorithms, SAL programs also help exploit knowledge of 
physical properties such as continuity and locality in data fields.
They work in a bottom-up manner to uncover regions of uniformity in
spatially distributed data. In conjunction with this process, we introduce
a top-down sampling strategy that focuses data collection in only those
regions that are deemed most important to support a data mining
objective. Together, they help define a methodology for mining in data-scarce 
domains. We describe
this methodology at a high-level and devote the major part of the paper to
two applications that employ it.

\section{A Methodology for Mining in Data-Scarce Domains}
It is possible to study the problem of sampling for targeted data mining activities, such
as clustering, finding association rules, and decision tree construction~\cite{ganti-ieee}. This is
the approach taken by work such as~\cite{mannila}. In this paper, however, we are interested in a 
general framework or language to express data mining operations on datasets and 
which can be used to study the design of data collection and sampling strategies. The spatial 
aggregation language (SAL)~\cite{bailey-kellogg96,yip96a} is such a framework. 

\subsection{SAL: The Spatial Aggregation Language}
As a data mining framework, SAL
is based on successive manipulations of data fields by a uniform vocabulary of
aggregation, classification, and abstraction operators. Programming in SAL follows
a philosophy of building a multi-layer hierarchy of aggregations of data. These
increasingly abstract descriptions of data are built using explicit representations
of physical knowledge, expressed as metrics, adjacency relations, and
equivalence predicates. This allows a SAL program to uncover and exploit structures in
physical data. 

SAL programs employ what has been called an {\em imagistic reasoning} style~\cite{yip95b}.
They employ vision-like routines to manipulate multi-layer geometric and
topological structures in spatially distributed data.  SAL adopts a
{\em field ontology}, in which the input is a {\em field} mapping from
one continuum to another (e.g.\ 2-D temperature field: $\R^2
\rightarrow \R^1$; 3-D fluid flow field: $\R^3 \rightarrow \R^3$).
Multi-layer structures arise from continuities in fields at multiple
scales.  Due to continuity, fields exhibit regions of uniformity, and
these regions of uniformity can be abstracted as higher-level
structures which in turn exhibit their own continuities.
Task-specific domain knowledge specifies how to uncover such regions
of uniformity, defining metrics for closeness of both field objects
and their features.  For example, isothermal contours are connected
curves of nearby points with equal (or similar enough) temperature.

The identification of structures in a field is a form of data
reduction: a relatively information-rich field representation is
abstracted into a more concise structural representation (e.g.\
pressure data points into isobar curves or pressure cells; isobar
curve segments into troughs).  Navigating the mapping from field to
abstract description through multiple layers rather than in one giant
step allows the construction of more modular programs with more
manageable pieces that can use similar processing techniques at
different levels of abstraction.  The multi-level mapping also allows
higher-level layers to use global properties of lower-level objects as
local properties of the higher-level objects.  For example, the
average temperature in a region is a global property when considered
with respect to the temperature data points, but a local property when
considered with respect to a more abstract region description.  As
this paper demonstrates, analysis of higher-level structures in such a
hierarchy can guide interpretation of lower-level data.

\begin{figure}
\begin{center}
\includegraphics[width=3.5in]{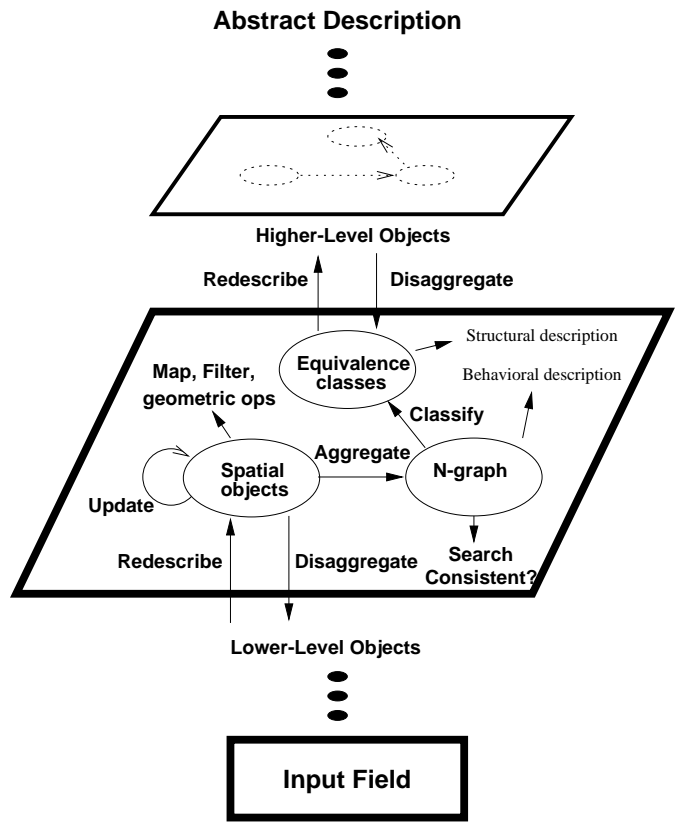}
\end{center}
\caption{SAL multi-layer spatial aggregates, uncovered by a uniform
vocabulary of operators utilizing domain knowledge. A variety of scientific data mining 
tasks, such as vector field bundling, contour aggregation, correspondence abstraction, clustering,
and
uncovering regions of uniformity can be expressed as multi-level computations with SAL
aggregates.}
\label{fig:sa}
\end{figure}
\begin{figure}
\begin{center}
\begin{tabular}{cccc}
\includegraphics[width=1.5in]{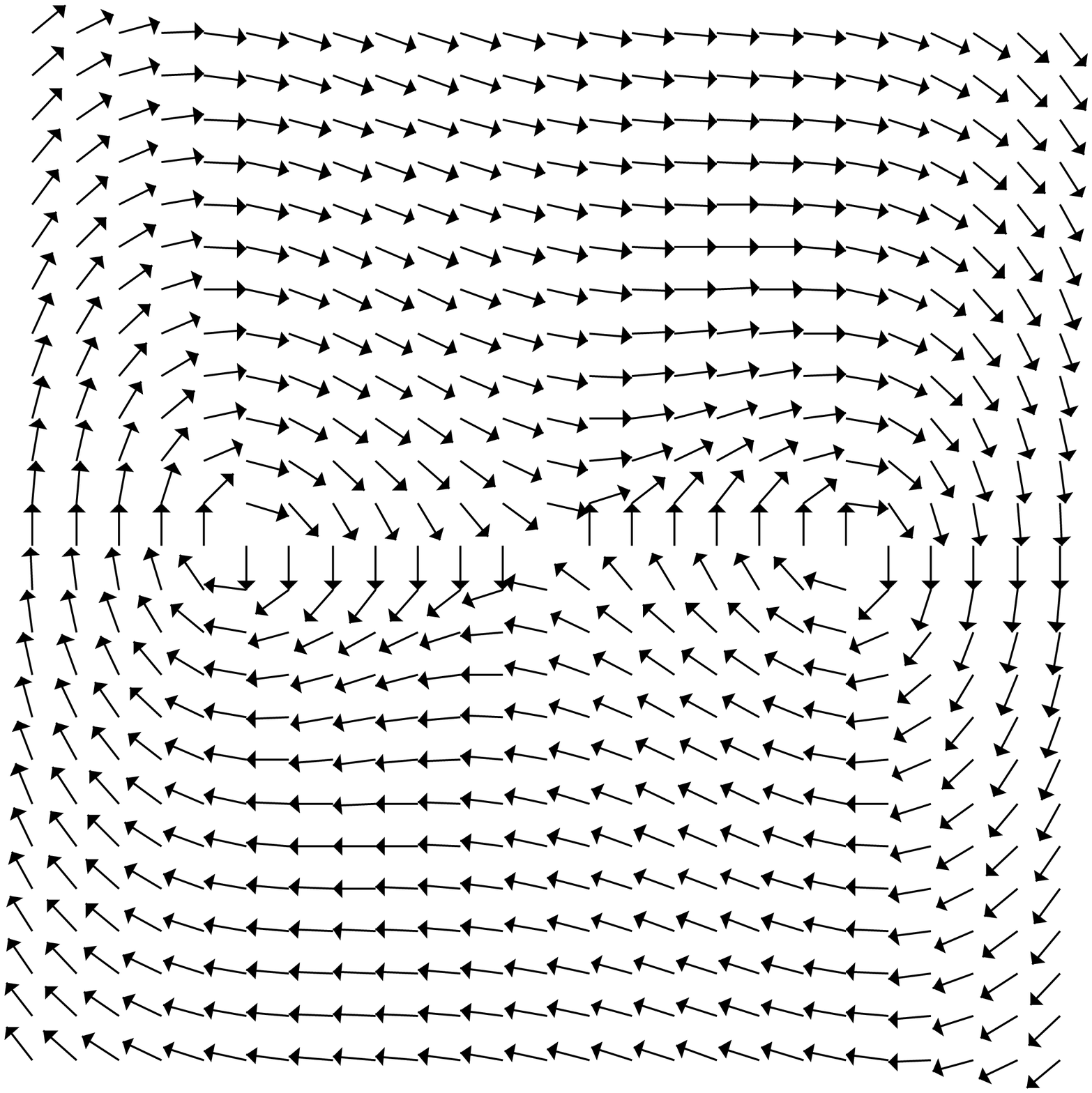} &
\includegraphics[width=1.5in]{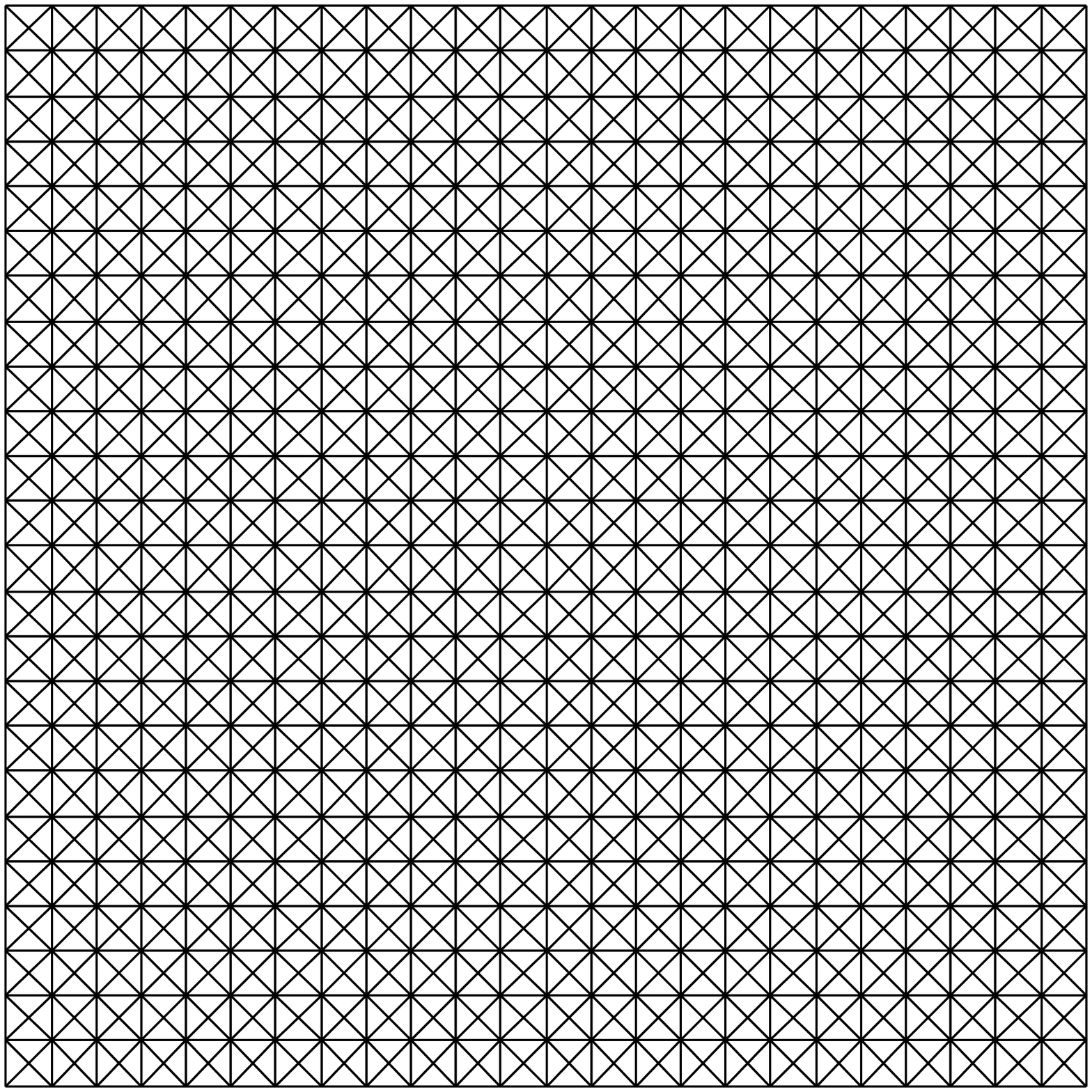} &
\includegraphics[width=1.5in]{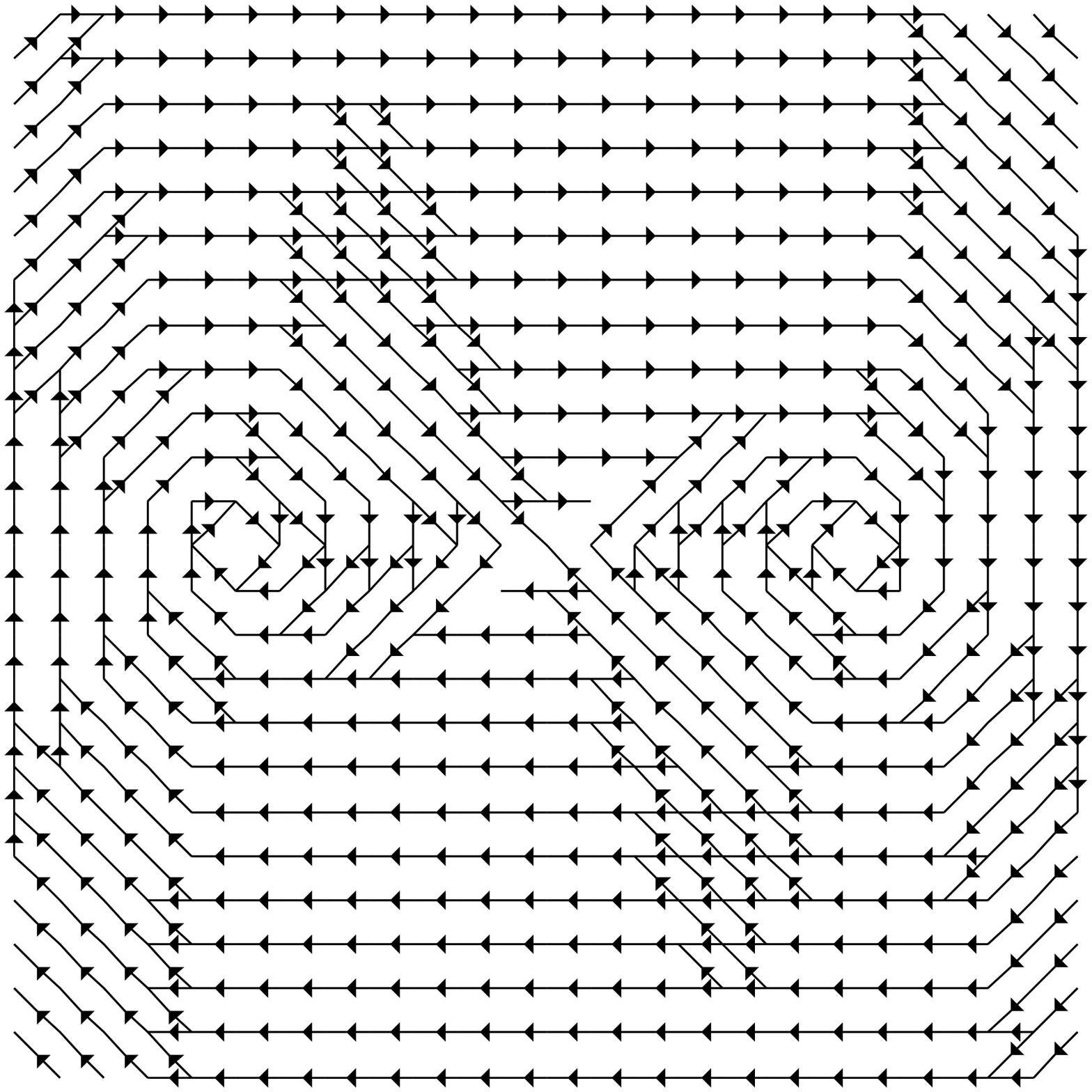} &
\includegraphics[width=1.5in]{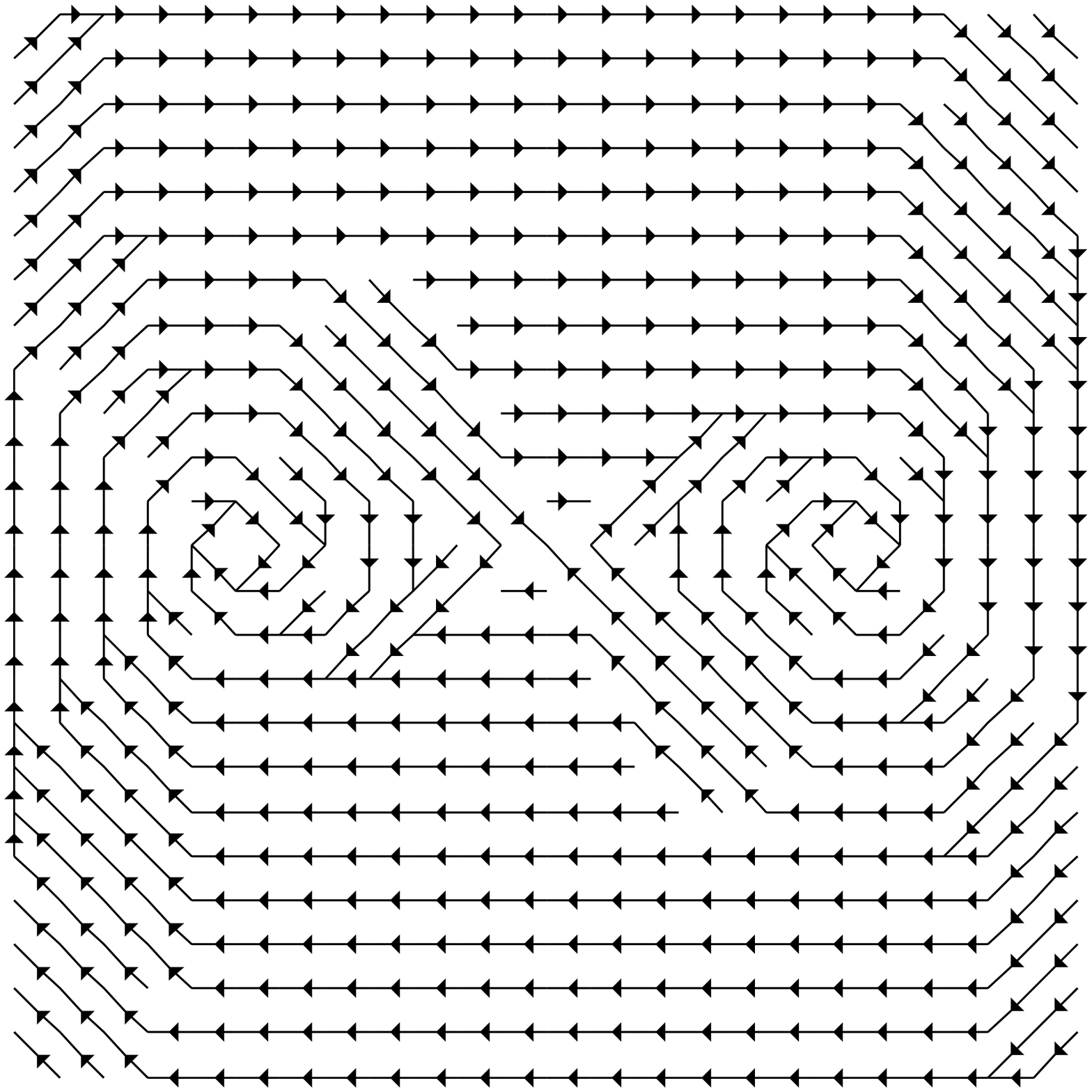} \\
(a) & (b) & (c) & (d) \\
\includegraphics[width=1.5in]{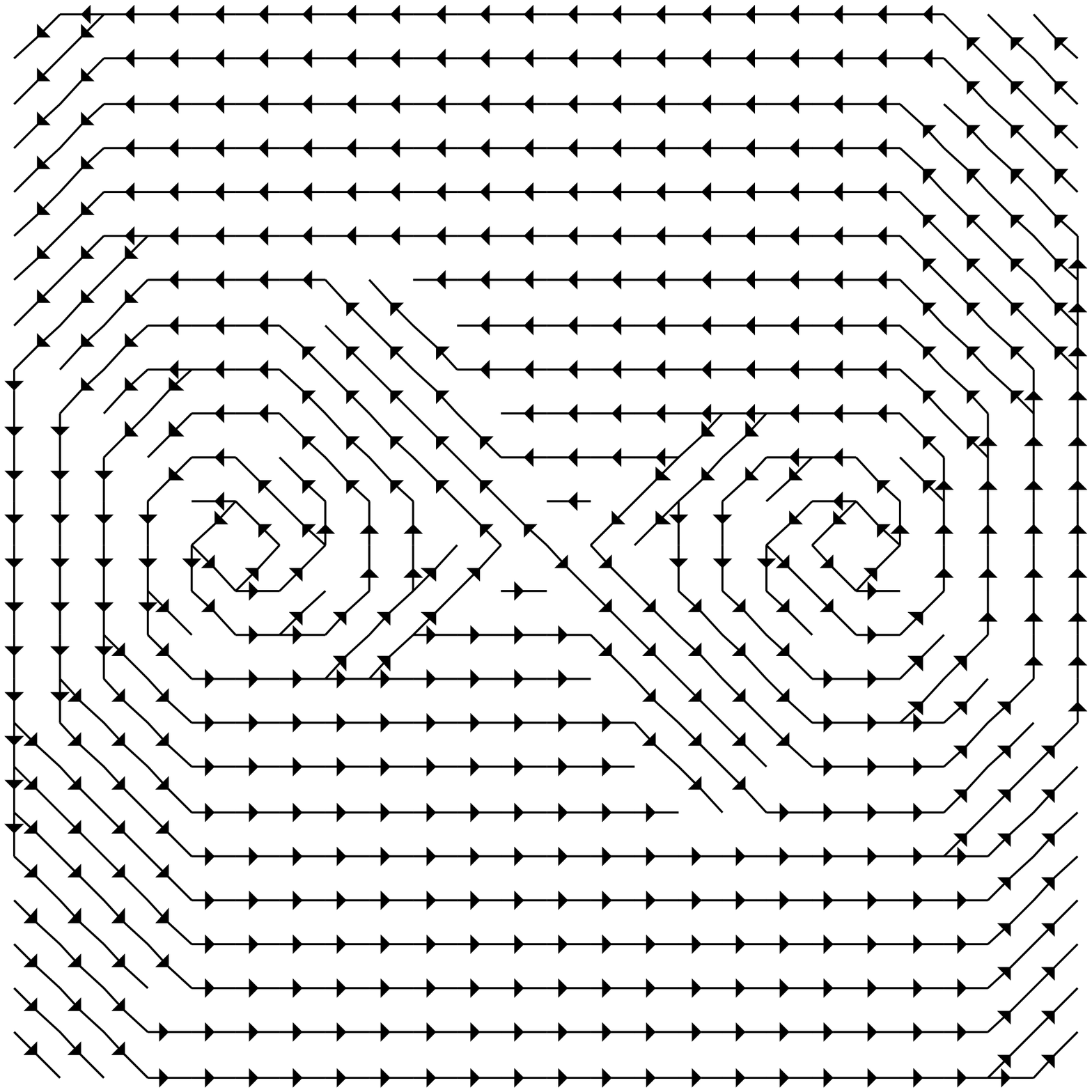} &
\includegraphics[width=1.5in]{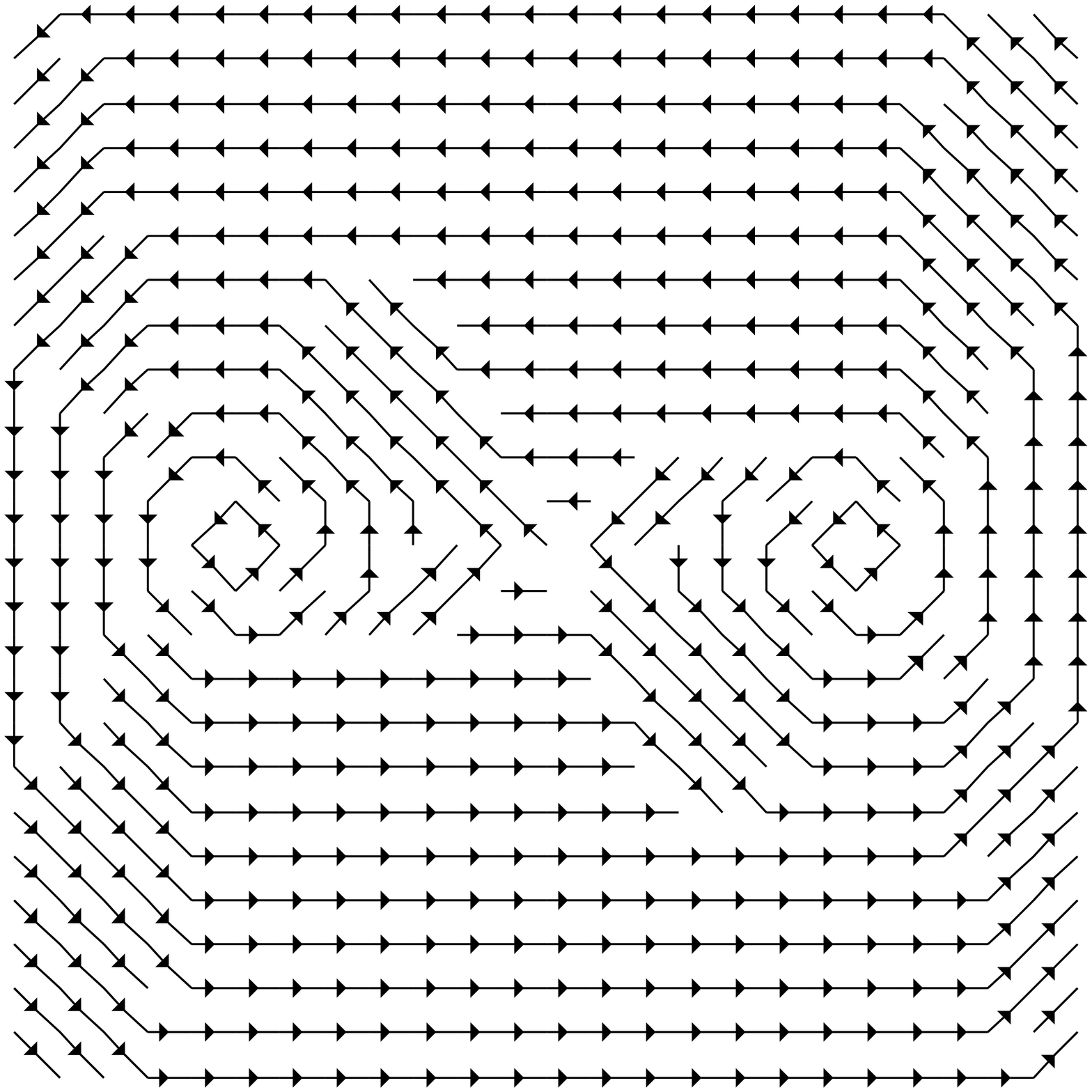} &
\includegraphics[width=1.5in]{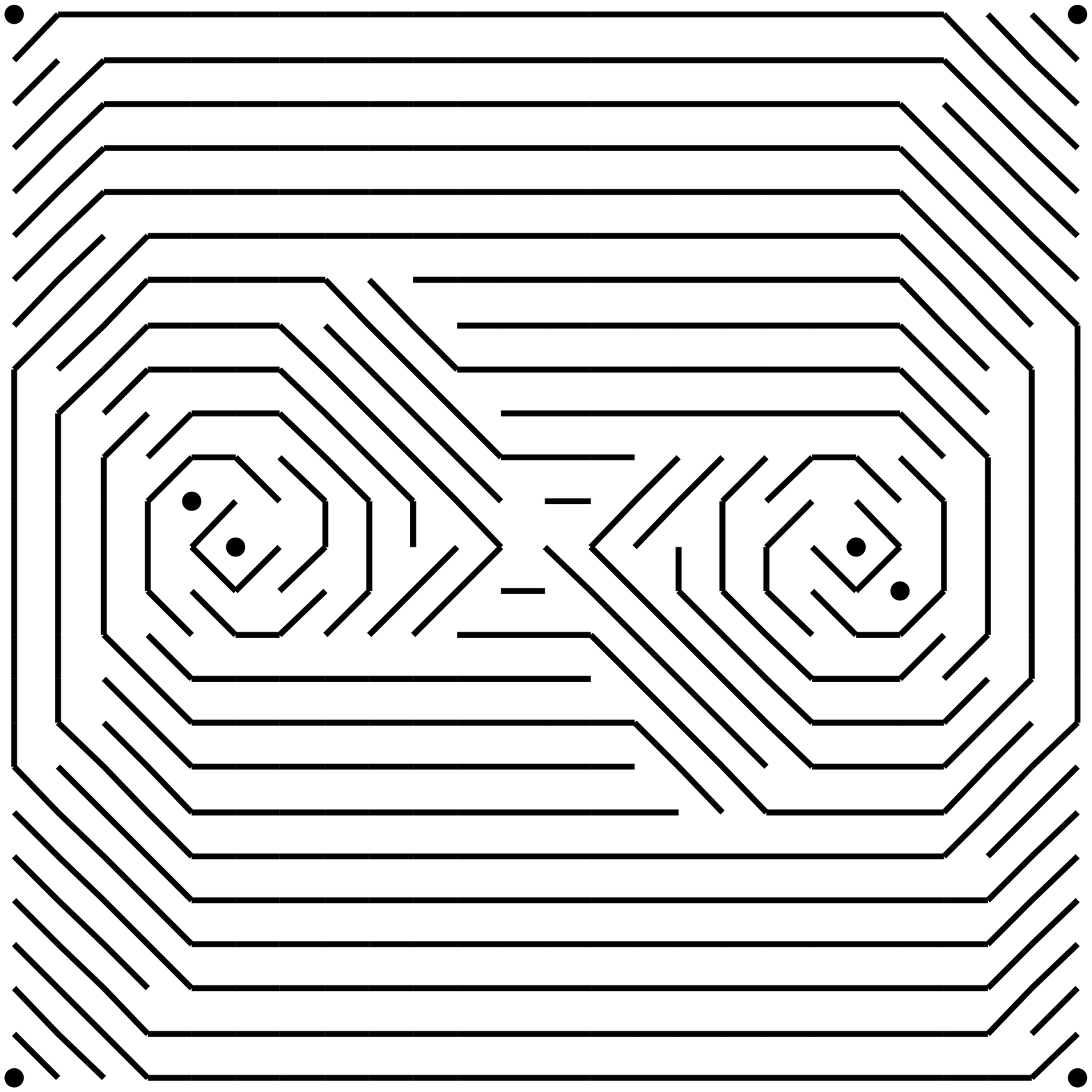} &
\includegraphics[width=1.5in]{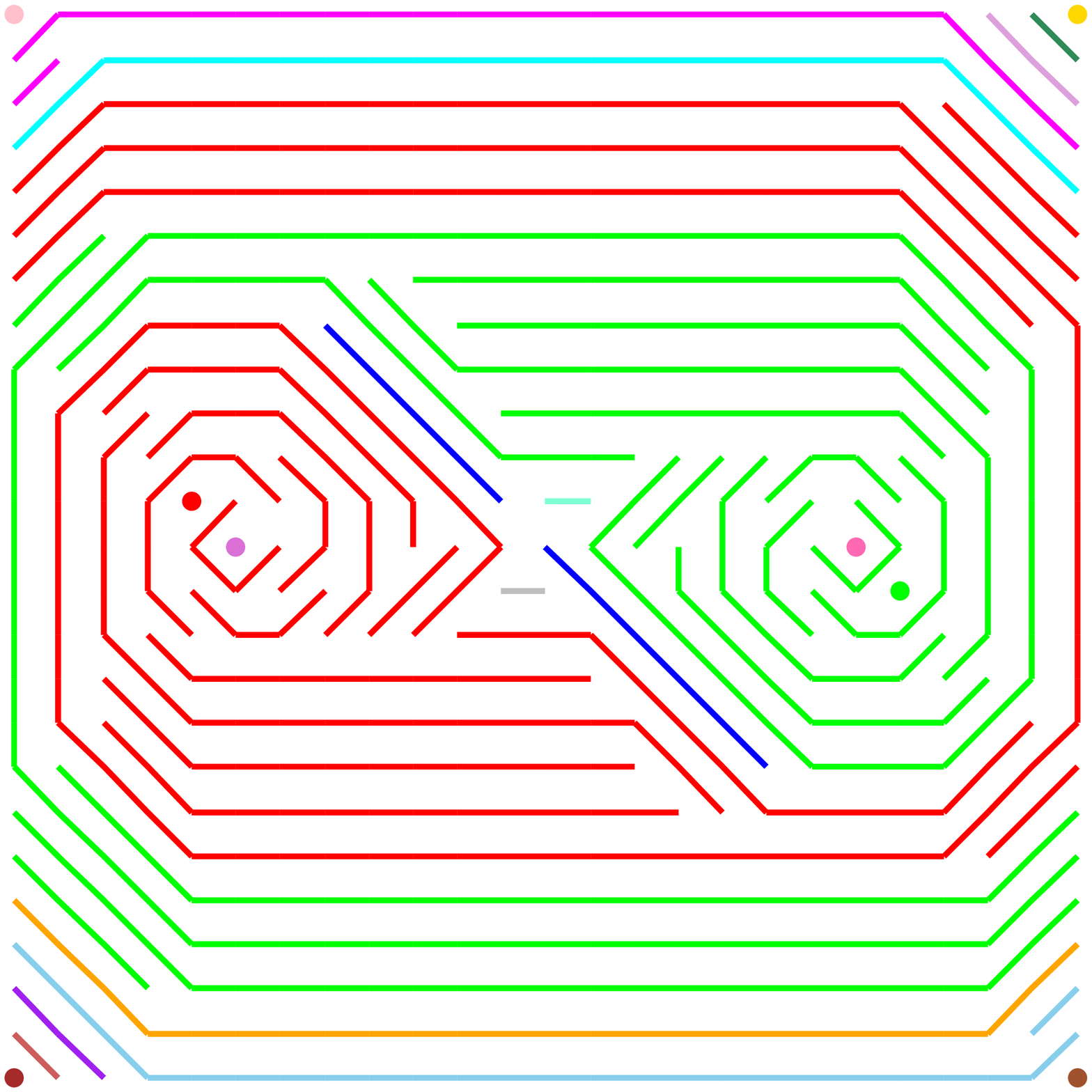} \\
(e) & (f) & (g) & (h) \\
\end{tabular}
\end{center}
\caption{Example steps in SAL implementation of vector field
analysis application.  (a) Input vector field. (b) 8-adjacency
neighborhood graph.  (c) Forward neighbors. (d) Best forward
neighbors. (e) Ngraph transposed from best forward neighbors. (f) Best
backward neighbors. (g) Resulting adjacencies redescribed as
curves. (h) Higher-level aggregation and classification of curves
whose flows converge.}
\label{fig:vect}
\end{figure}

\begin{figure}
\begin{tabular}{|lc|} \hline
 & \\
\begin{minipage}{\textwidth}
\small
\begin{alltt}
// (a) Read vector field.
vect_field = read_point_point_field(\emph{infile});
points = domain_space(vect_field);

// (b) Aggregate with 8-adjacency (i.e. within 1.5 units).
point_ngraph = aggregate(points, make_ngraph_near(1.5));

// (c) Compare vector directions with node-neighbor direction.
angle = function (p1, p2) \{
  dot(normalize(mean(feature(vect_field, p1), feature(vect_field, p2))),
      normalize(subtract(p2, p1)))
\}
forward_ngraph = filter_ngraph(adj in point_ngraph, \{
 angle(from(adj), to(adj)) > \emph{angle\_similarity}
\})
// (d) Find best forward neighbor, comparing vector direction 
// with ngraph edge direction and penalizing for distance.
forward_metric = function (adj) \{
  angle(from(adj), to(adj)) - \emph{distance\_penalty} * distance(from(adj),to(adj))
\}
best_forward_ngraph = best_neighbors_ngraph(forward_ngraph, forward_metric);

// (e) Find backward neighbors by transposing best forward neighbors.
backward_ngraph = transpose_ngraph(best_forward_ngraph);

// (f) At junctions, keep best backward neighbor using metric
// similar to that for best forward neighbors.
backward_metric = function (adj) \{
  angle(to(adj), from(adj)) - \emph{distance\_penalty}*distance(from(adj),to(adj))
\}
best_backward_ngraph = best_neighbors_ngraph(backward_ngraph, backward_metric);

// (g) Move to a higher abstraction level by forming equivalence classes
// from remaining groups and redescribing them as curves.
final_ngraph = symmetric_ngraph(best_backward_ngraph, extend=true);
point_classes = classify(points, make_classifier_transitive(final_ngraph));

points_to_curves = redescribe(classes(point_classes),
   make_redescribe_op_path_nline(final_ngraph));
trajs = high_level_objects(points_to_curves);
\end{alltt}
\end{minipage} 
 & \\
 & \\
\hline
\end{tabular}
\caption{SAL data mining program for the vector field analysis application of Fig.~\ref{fig:vect}.} 
\label{samplecode}
\end{figure}

SAL supports structure discovery through a small set of generic
operators, parameterized with domain-specific knowledge, on uniform
data types.  These operators and data types mediate increasingly
abstract descriptions of the input data (see Fig.~\ref{fig:sa}) to
form higher-level abstractions and mine patterns.  The {\em
primitives} in SAL are contiguous regions of space called {\em spatial
objects}; the {\em compounds} are (possibly structured) collections of
spatial objects; the {\em abstraction mechanisms} connect collections
at one level of abstraction with single objects at a higher level.

SAL is currently available as a C++ library\footnote{The SAL implementation can be
downloaded from http://www.cis.ohio-state.edu/insight/sal-code.html.} providing access to a
large set of data type implementations and operations.  In addition,
an interpreted, interaction environment layered over the library
supports rapid prototyping of data mining applications.  It allows
users to inspect data and structures, test the effects of different
predicates, and graphically interact with representations of the
structures.  

To illustrate SAL programming style, consider the task of bundling 
vectors in a given vector field (e.g.\ wind velocity or temperature gradient)
into a set of streamlines (paths through the field following the
vector directions). This process can be depicted as shown in
Fig.~\ref{fig:vect} and the corresponding SAL data mining program is shown
in Fig.~\ref{samplecode}.
The steps
in this program are as follows:
(a) Establish a {\em field} mapping points (locations) to points
(vector directions, assumed here to be normalized).  (b) Localize
computation with a {\em neighborhood graph}, so that only spatially
proximate points are compared.  
(c)--(f) Use a series of local computations on this representation to
find {\em equivalence classes} of neighboring vectors with respect to
vector direction (systematically eliminate all edges but those whose
directions best match the vector direction at both endpoints).
(g) {\em Redescribe} equivalence classes of vectors into more abstract
streamline curves.  (h) Aggregate and classify these curves into
groups with similar flow behavior, {\em using the exact same operators
but with different metrics} (code not shown).  As this example
illustrates, SAL provides a vocabulary for expressing the knowledge
required (e.g., distance metrics and similarity metrics) 
for uncovering multi-level structures in spatial datasets.  It has been
applied to applications ranging from decentralized control
design~\cite{bailey-kellogg01}
to analysis of diffusion-reaction morphogenesis~\cite{ordonez00}.

\subsection{Data Collection and Sampling}
\begin{figure}
\begin{center}
\includegraphics[height=3in]{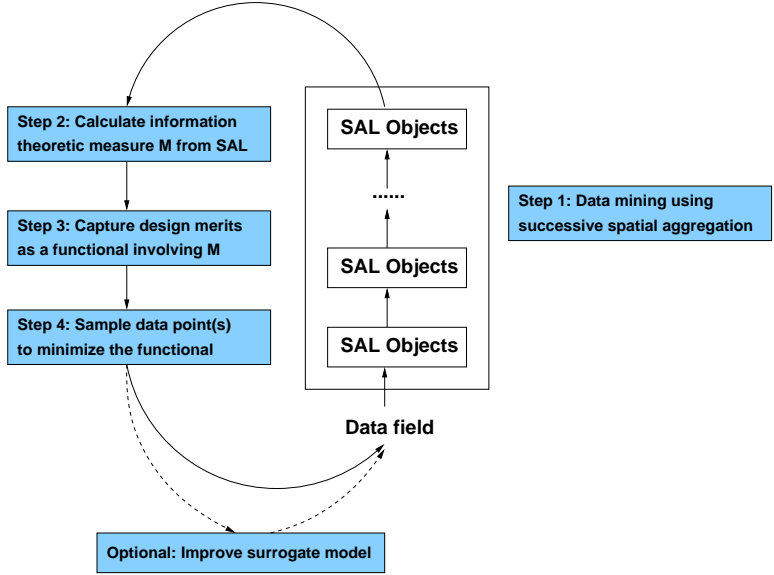}
\end{center}
\caption{The sampling methodology for SAL mining in data-scarce domains.}
\label{sampling-meth}
\end{figure}

The above example illustrated the use of SAL in a data-rich domain. The exploitation 
of physical properties is a central tenet of SAL since it drives the computation of
multi-level spatial aggregates. Many important physical properties can be expressed as
SAL computations by suitably defining adjacency relations and aggregation metrics.
To extend the use of SAL to data-scarce settings, we
present the sampling methodology outlined in Fig.~\ref{sampling-meth}.

Once again, it is easy to understand the methodology in the context of the vector-field bundling
application (Fig.~\ref{fig:vect}). Assume that we apply the SAL data mining program of Fig.~\ref{samplecode}
with a small dataset and have navigated upto the highest level of the hierarchy (streamlines bundled with
convergent flows). 
The SAL program computes different streamline aggregations from a neighborhood graph and chooses
one based on how well its curvature matches the direction of the vectors it aggregates. If data
is scarce, it is likely that some of these classification decisions will be {\it ambiguous}, i.e.,
there may exist multiple streamline aggregations. {\bf In such a case, we would like to choose a new data sample
that reduces the ambiguity and clarifies what the correct classification should be.} 

This is the essence of our sampling methodology: using SAL aggregates, we identify an information-theoretic measure
(here, ambiguity) that can be used to drive stages of future data collection. For instance, the
ambiguous streamline classifications can be summarized as a 2D ambiguity distribution that has a spike
for every location where an ambiguity was detected.
Reduction of ambiguity can be posed as the problem of minimization of (or maximization, as the case may be) 
a functional involving the (computed) ambiguity. The functional could be the entropy in the underlying 
data field, as revealed by the ambiguity distribution.
Such a minimization will lead us to selecting a data point(s) that clarifies the distribution of
streamlines, and hence makes more effective use of data for data mining purposes.  The net effect of this methodology is 
that we are able to capture the desirability of a particular design (data layout) in terms 
of computations involving SAL aggregates. Thus, sampling is conducted for the express purpose of improving the
quality and efficacy of data mining. The dataset is updated with the newly collected value and the process is repeated
till a desired stopping criteria is met. For instance, we could terminate if the 
functional is within accepted bounds, or
when there is no improvement in confidence of data mining results between successive rounds of data collection.
In our case, when there is no further ambiguity.

This idea of sampling to satisfy particular design criteria has been studied in various
contexts, especially spatial statistics~\cite{Easterling, journel, dace}. 
Many of these approaches (including ours) rely on
capturing properties of a desirable design in terms of a novel
objective function.  The distinguishing feature of our work is that it
uses {\em spatial} information gleaned from a higher level of
abstraction to focus data collection at the field/simulation code
layer. 
The applications presented here are also novel in that they span and connect
arbitrary levels of abstraction, thus suggesting new ways to integrate
qualitative and quantitative simulation~\cite{berleant-kuipers}.

%

We present concrete realizations of the above methodology in the next section. 
But before we proceed, it is
pertinent to note an optional step in our methodology. The newly collected data value can be used to improve a {\it
surrogate} model which then generates a dense data field for mining.
A surrogate function is something that is used in lieu of the real
data source, so as to generate sufficient data for mining purposes. This is often
more advantageous than working directly with sparse data. Surrogate models are widely used in
engineering design, optimization, and in response surface approximations~\cite{ltw,response-book}.

%

Together, SAL and our focused sampling methodology address the main issues raised in
the beginning of the paper: SAL's uniform use of fields and abstraction operators allows
us to exploit prior knowledge in a bottom-up manner. Discrepancies as suggested by our
knowledge of physical properties (e.g., ambiguities) are used in a top-down manner by
the sampling methodology. Continuing these two stages alternatively leads to a closed-loop
data mining solution for data-scarce domains.

\section{Example Applications}
\subsection{Mining Pockets in Spatial Data}
Our first application is motivated by the aircraft design problem and is meant 
to illustrate the basic idea of our methodology. Here, we are given a spatial vector field
and we wish to identify {\it pockets} underlying the gradient. In a weather map, this might
mean identifying pressure troughs, for instance. The question is: `where should data be
collected so that we are able to mine the pockets with high confidence?' We begin by presenting 
a mathematical function that gives rise to pockets in spatial fields. This function will 
be used to validate and test our data mining and sampling methodology.

\begin{figure}
\begin{center}
\begin{tabular}{cc}
\includegraphics[width=2.5in]{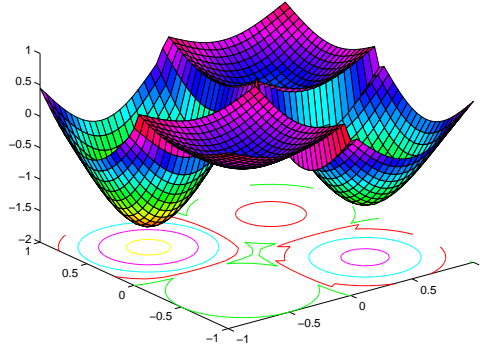} 
\end{tabular}
\end{center}
\caption{A 2D pocket function.}
\label{fig:pocket-diag}
\end{figure}

\subsubsection*{de Boor's function}
Carl de Boor invented a pocket function that exploits containment properties of the
$n$-sphere of radius 1 centered at the origin ($\Sigma {x_i}^2 \leq 1$) with respect to the 
$n$-dimensional hypercube defined by $x_i \in [-1, 1], i=1\cdots n$. Even though the 
sphere is embedded inside the cube, notice that the ratio of the volume of the cube ($2^n$) to that of the sphere
($\pi^{n/2} / (n/2)!$) grows unboundedly with $n$. This means that the
volume of a high-dimensional cube is concentrated in its corners (a
counterintuitive notion at first). de Boor exploited this
property to design a difficult-to-optimize function which assumes a
{\it pocket} in each corner of the cube (Fig.~\ref{fig:pocket-diag}), that
is just outside the sphere.  Formally, it can be
defined as:
\begin{eqnarray}
\alpha({\mathbf X}) & = & cos \left( \sum_{i=1}^n 2^i \left( 1 + {x_i \over{\mid x_i \mid}}\right) \right) - 2 \\
\delta({\mathbf X})     & = & \| {\mathbf{X}} - 0.5 {\mathbf{I}}\| \\
p({\mathbf X}) & = & \alpha({\mathbf X}) ( 1 - \delta^2({\mathbf X})
(3 - 2\delta({\mathbf X}))) + 1
\end{eqnarray}
where ${\mathbf X}$ is the n-dimensional point $(x_1,x_2,\cdots,x_n)$
at which the pocket function $p$ is evaluated, ${\mathbf I}$ is the
identity n-vector, and $\|\cdot\|$ is the $L_2$ norm.

It is easily seen that $p$ has $2^n$ pockets (local minima); if $n$ is large
(say, 30, which means it will take more than half a million points to
just represent the corners of the $n$-cube!), naive global
optimization algorithms will require an unreasonable number of
function evaluations to find the pockets. Our goal for data mining here is to obtain a 
qualitative indication of the existence, number, and locations of pockets, using 
low-fidelity models and/or as few data points as possible.  The results can then be
used to seed higher-fidelity calculations.  This is also fundamentally
different from DACE~\cite{dace}, polynomial response surface
approximations~\cite{ltw}, and other approaches in geo-statistics
where the goal is accuracy of functional prediction at untested data
points.  Here, accuracy of estimation is traded for the ability to
mine pockets.

\subsubsection*{Surrogate Function}
In this study, we use the SAL vector-field bundling code presented earlier along with
a surrogate model as the basis for generating a dense field
of data. Surrogate theory is an established area in engineering optimization and
there are several ways in which we can build a surrogate.
However, the local nature of SAL computations means that we can be selective about
our choice of surrogate representation.
For example, global, least-squares type
approximations are inappropriate since measurements at all locations
are equally considered to uncover trends and patterns in a particular
region.  We advocate the use of kriging-type
interpolators~\cite{dace}, which are local modeling methods with roots
in Bayesian statistics.  Kriging can handle situations with multiple
local extrema (for example, in weather data, remote sensing data,
etc.) and can easily exploit anisotropies and trends. Given $k$
observations, the interpolated model gives exact responses at these
$k$ sites and estimates values at other sites by minimizing the mean
squared error (MSE), assuming a random data process with zero mean and
a known covariance function.

Formally (for two dimensions), the true function $p$ is assumed to be
the realization of a random process such as:
\begin{equation}
p(x,y) = \beta + Z(x,y)
\end{equation}
where $\beta$ is typically a uniform random variate, estimated based
on the known $k$ values of $p$, and $Z$ is a correlation function.
Kriging then estimates a model $p'$ of the same form, based on the
$k$ observations:
\begin{equation}
p'(x_i,y_i) = E(p(x_i,y_i) \midv p(x_1,y_1), \cdots, p(x_k,y_k))
\end{equation}
and minimizing mean squared error between $p'$ and $p$:
\begin{equation}\label{eq:MSE} 
MSE = E(p'(x,y) - p(x,y))^2
\end{equation}
A typical choice for $Z$ in $p'$ is $\sigma^2 R$, where scalar
$\sigma^2$ is the {\it estimated} variance, and correlation matrix $R$
encodes domain-specific constraints and reflects the current fidelity
of data.  We use an exponential function for entries in $R$, with two
parameters $C_1$ and $C_2$:
\begin{equation}\label{eq:R} 
R_{ij} = e^{-C_1|x_i-x_j|^2 - C_2|y_i-y_j|^2}
\end{equation} 
Intuitively, values at closer points should be more highly correlated.

The estimator minimizing mean squared error is then obtained by 
multi-dimensional optimization (the derivation from Eqs.~\ref{eq:MSE}
and~\ref{eq:R} is beyond the scope of this paper):
\begin{equation}\label{eq:optim1}
\max_C {\frac{-k}{2}}(\ln\sigma^2 + \ln |R|)
\end{equation}
This expression satisfies the conditions that there is no error
between the model and the true values at the chosen $k$ sites, and
that all variability in the model arises from the design of $Z$.  The
multi-dimensional optimization is often performed by gradient descent
or pattern search methods.  More details are available in~\cite{dace},
which demonstrates this methodology in the context of the design and
analysis of computer experiments.

\subsubsection*{Data Mining and Sampling Methodology}
The bottom-up computation of SAL aggregates from the surrogate model's outputs
will possibly lead to some ambiguous streamline classifications, as discussed earlier. 
Ambiguity can reflect the desirability of acquiring data at or near a
specified point, to clarify the correct classification and to serve as
a mathematical criterion of information content.
There are several ways in which we can use information about ambiguity to drive
data collection. In this study, we express the ambiguities as a distribution describing
the number of possible good neighbors (for a streamline).
This {\it ambiguity distribution} provides a novel mechanism to include
qualitative information --- streamlines that agree will generally
contribute less to data mining, for information purposes. The information-theoretic measure
$M$ (ref. Fig.~\ref{sampling-meth}) was thus defined to be the ambiguity distribution $\wp$.

The functional was defined as the posterior entropy $E(-\log d)$, where $d$ is the conditional
density of $\wp$ over the design space {\it not covered} 
by the current data values. By a reduction argument, minimizing this posterior entropy can be
shown to be maximizing the prior entropy over the {\it unsampled} design space~\cite{dace}.
In turn, this means that the amount of information obtained from an experiment (additional data
collection) is maximized. In addition, we also incorporated $\wp$ as an indicator covariance term in
our surrogate model (this is a conventional method
for including qualitative information in an interpolatory model~\cite{journel}). 

\subsubsection*{Experimental Results}
The initial experimental configuration used a face-centered design ($4$ points in the 2D case).  A 
surrogate model by kriging interpolation then generated data on a $41^n$-point grid. 
de Boor's function was used as the source for data values; we also employed pseudorandom perturbations
of it that shift the pockets from the corners in a somewhat unpredictable
way (see~\cite{ambig} for details). In total, we experimented with 100 perturbed
variations (each) of the 2D and 3D pocket functions. For each of these cases, data collection was organized
in rounds of one extra sample each (that minimizes the above functional). The number of samples needed
to mine all the pockets by SAL was recorded. We also compared our results with those obtained 
from a pure DACE/kriging approach (i.e., where sampling was directed at improving accuracy of function estimation). In other words, we used the DACE methodology to suggest
new locations for data collection and determined how these choices fared with respect
to mining the pockets.

\begin{figure}
\begin{center}
\includegraphics[width=3in]{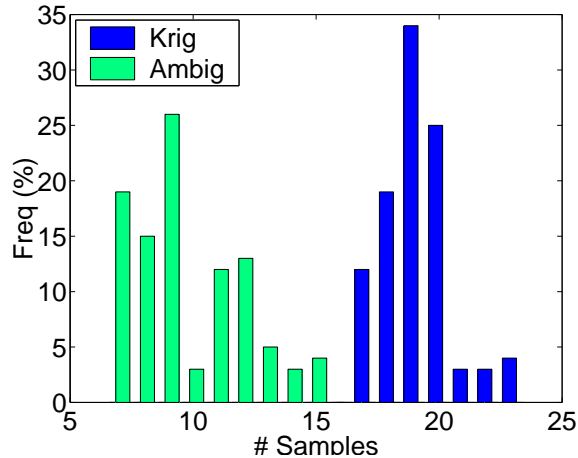}
\end{center}
\caption{Pocket-finding results (2D) show that focused sampling using a measure
of ambiguity always requires fewer total samples (7-15) than conventional kriging (17-23).}
\label{fig:pocket-bar}
\end{figure}

\begin{figure}
\begin{center}
\begin{tabular}{cc}
\includegraphics[width=3in]{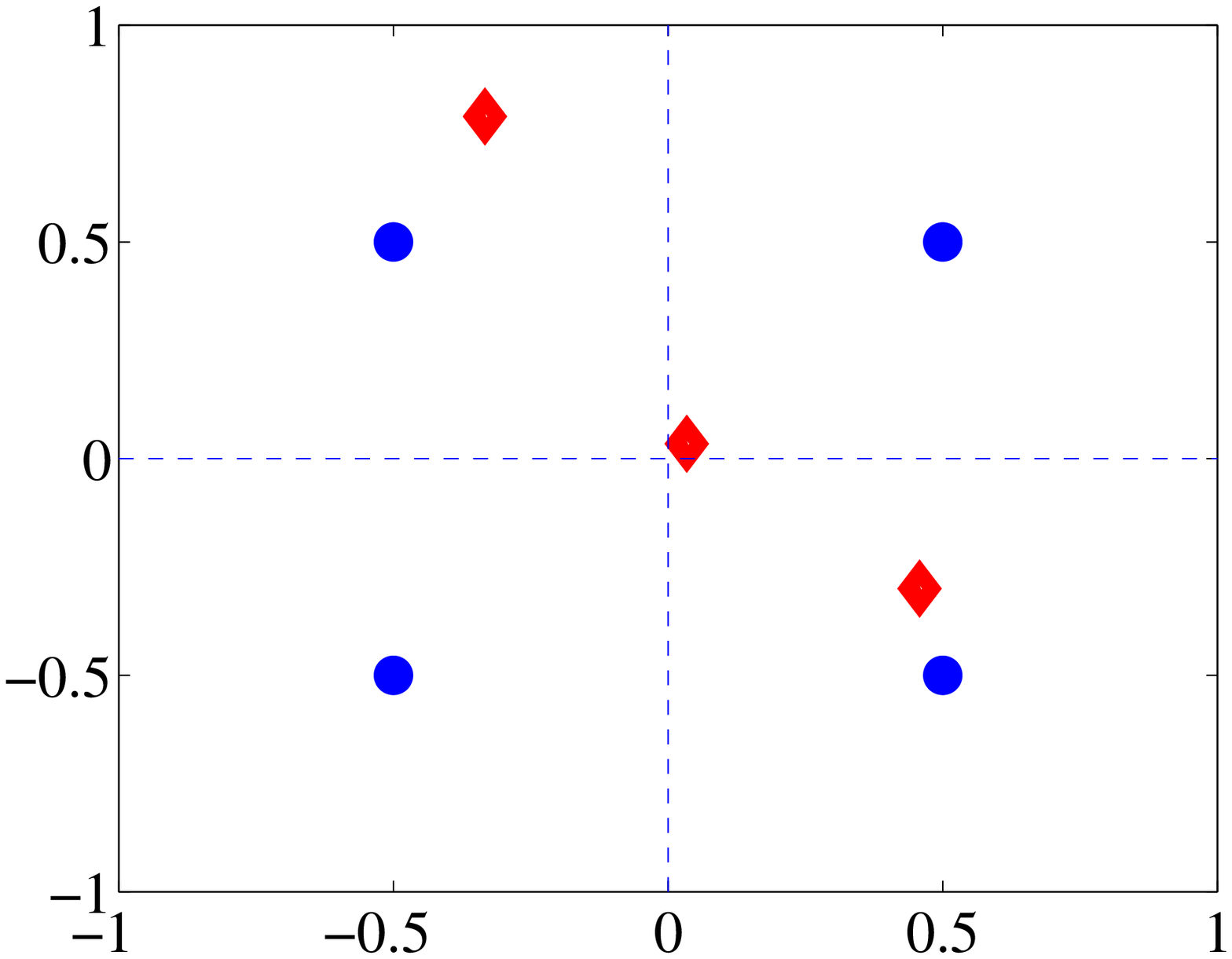} &
\includegraphics[width=2.5in]{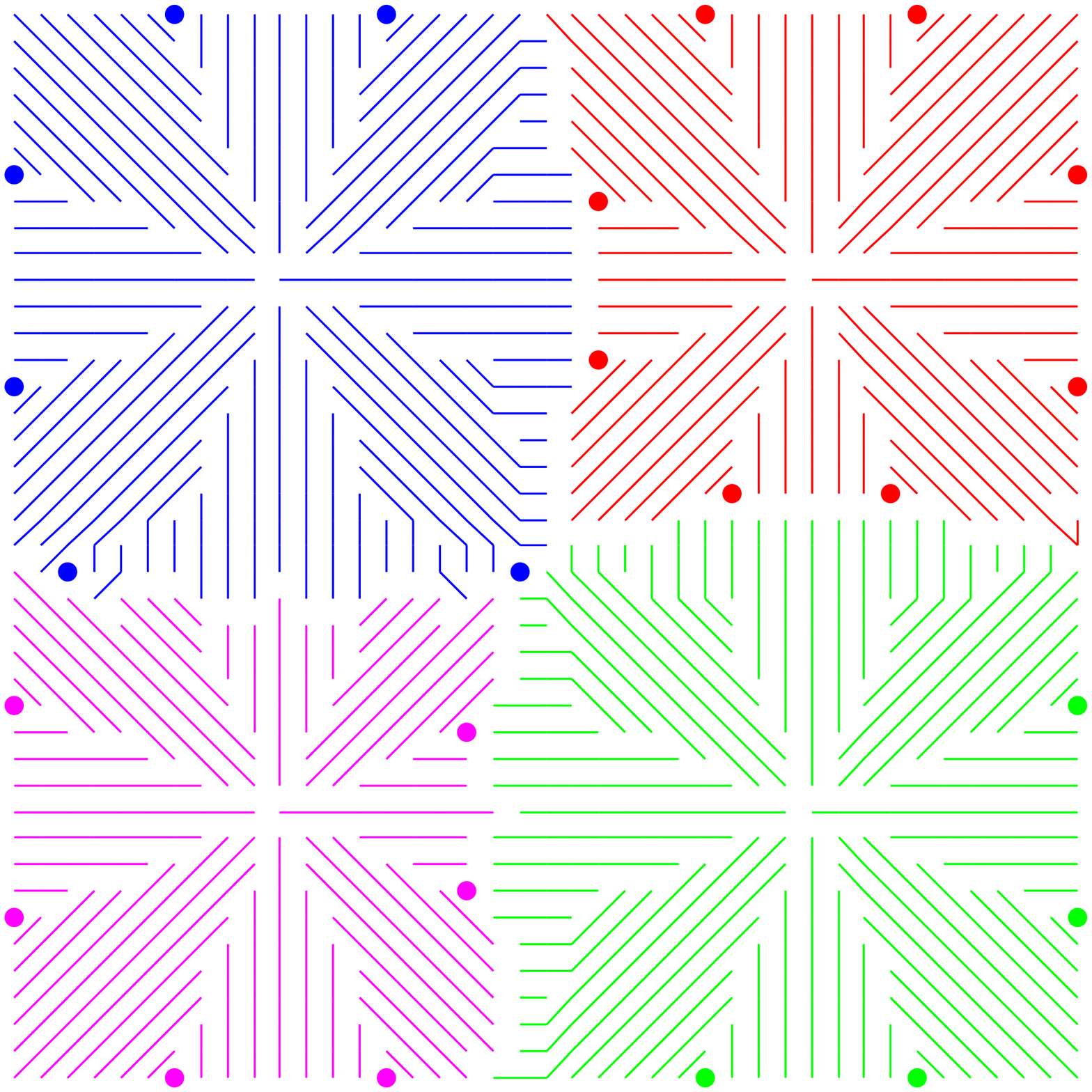} 
\end{tabular}
\end{center}
\caption{Mining pockets in 2D from only 7 sample points. 
(left) 
The chosen sample locations: 4 initial
face-centered samples (marked as blue circles) plus 3 
samples selected by our methodology (marked as red diamonds). Note that no additional sample is required in
the lower-left quadrant.  (right)
SAL structures in surrogate
model data, confirming the existence of four pockets.}
\label{fig:pocket}
\end{figure}

Fig.~\ref{fig:pocket-bar} shows the distributions of total number of data samples
required to mine the four pockets for the 2D case. We were thus able to mine the 2D pockets
using 3 to 11 additional samples, whereas the conventional kriging approach required
13 to 19 additional samples. The results were were more striking in the 3D case:
at most 42 additional samples for focused sampling and upto 151 points for conventional
kriging. This shows that our focused sampling methodology performs 40-75\% better
than sampling by conventional kriging.

Fig.~\ref{fig:pocket} (left) 
describes a 2D design involving only $7$ total data points that is able to mine the four pockets.
Counterintuitively, no additional sample is required in the lower left quadrant! While this
will lead to a highly sub-optimal design (from the traditional viewpoint
of minimizing variance in predicted values), it is nevertheless an appropriate design
for data mining purposes. In particular, this means that neighborhood
calculations involving the other three quadrants are enough to uncover
the pocket in the fourth quadrant.  Since the kriging interpolator
uses local modeling and since pockets in 2D effectively occupy the
quadrants, obtaining measurements at ambiguous locations serves to
capture the relatively narrow regime of each dip, which in turn helps
to distinguish the pocket in the neighboring quadrant. This effect is hard
to achieve without exploiting knowledge of physical properties, in this case,
locality of the dips. 

\subsection{Qualitative Jordan Form Determination}
In our second  application, we use our methodology to identify the most probable
Jordan form of a given matrix. This is a good application for data mining
since the direct computation of the Jordan form leads to a numerically
unstable algorithm.

\subsubsection*{Jordan forms}
A matrix $\mathcal{A}$ (real or complex) that has $r$ independent eigenvectors has a Jordan form that
consists of $r$ {\it blocks}. Each of these blocks is an upper triangular
matrix that is associated with one of the eigenvectors of
$\mathcal{A}$ and whose size describes the multiplicity of the
corresponding eigenvalue. For the given matrix $\mathcal{A}$,
the diagonalization thus posits a nonsingular matrix $\mathcal{B}$ such that:
\begin{equation}
\mathcal{B}^{-1} \mathcal{A} \mathcal{B} = \left[ \begin{array}{cccc}
            {\mathcal{J}}_1 & & & \\
            & {\mathcal{J}}_2 & & \\
            & & \cdot & \\
            & & & {\mathcal{J}}_r\\
            \end{array}
\right]
\end{equation}
where
\begin{equation}
{\mathcal{J}}_i = \left[ \begin{array}{cccc}
                          \lambda_i & 1     &   & \\
                                    & \cdot & 1 & \\
                                    &       & \cdot  & 1\\
                                    &       &   & \lambda_i\\
\end{array}
\right]
\end{equation}
and $\lambda_i$ is the eigenvalue revealed by the $i$th Jordan block ($\mathcal{J}_i$).
The Jordan form is most easily explained by looking at how eigenvectors are
distributed for a given eigenvalue. Consider, for example, the matrix
$$ \left[ \begin{array}{crr}
1 & 1 & -1 \\
0 & 0 & 2 \\
0 & -1 & 3 \\
\end{array}
\right]$$
that has eigenvalues at 1, 1, and 2. This matrix has only two
eigenvectors, as revealed by the two-block structure of its Jordan form:
$$\left[ \begin{array}{cr|r}
1 & 1 & 0 \\
0 & 1 & 0 \\ \hline
0 & 0 & 2 \\
\end{array}
\right]$$
The Jordan form is unique modulo shufflings of the blocks and, in this case,
shows that there is one eigenvalue ($1$) of multiplicity $2$ and one eigenvalue
($2$) of multiplicty $1$. We say that the matrix has the
Jordan structure given by
$(1)^2 (2)^1$. In contrast, the matrix
$$ \left[ \begin{array}{crr}
1 & 0 & 0 \\
0 & 2 & 0 \\ 
0 & 0 & 1 \\
\end{array}
\right]$$
has the same eigenvalues but a three-block Jordan structure:
$$\left[ \begin{array}{c|r|r}
1 & 0 & 0 \\ \hline
0 & 1 & 0 \\ \hline
0 & 0 & 2 \\
\end{array}
\right]$$
This is because there are three independent eigenvectors (the unit vectors,
actually). The diagonalizing matrix is thus the identity matrix and the
Jordan form has three permutations. The Jordan structure is therefore
given by $(1)^1
(1)^1 (2)^1$. These two examples show that
a given eigenvalue's multiplicity could be distributed across one, many, or
all Jordan blocks. Correlating the eigenvalue with the block structure is
an important problem in numerical analysis.

The typical approach to computing the Jordan form is to `follow the staircase'
pattern of the structure and perform rank determinations in conjunction
with ascertaining the eigenvalues. One of the more serious
caveats with such an approach involves mistaking an eigenvalue of multiplicity
$> 1$ for multiple eigenvalues~\cite{staircase}. 
In the first example matrix
above, this might lead to inferring that the Jordan form has three
blocks. 
The extra care needed to safeguard staircase algorithms usually
involves more complexity than the original computation to be performed!
The ill-conditioned nature of this computation has thus
traditionally prompted numerical analysts to favor other, more stable,
decompositions.

\begin{figure}
\begin{center}
\begin{tabular}{cc}
\includegraphics[width=0.25\textwidth]{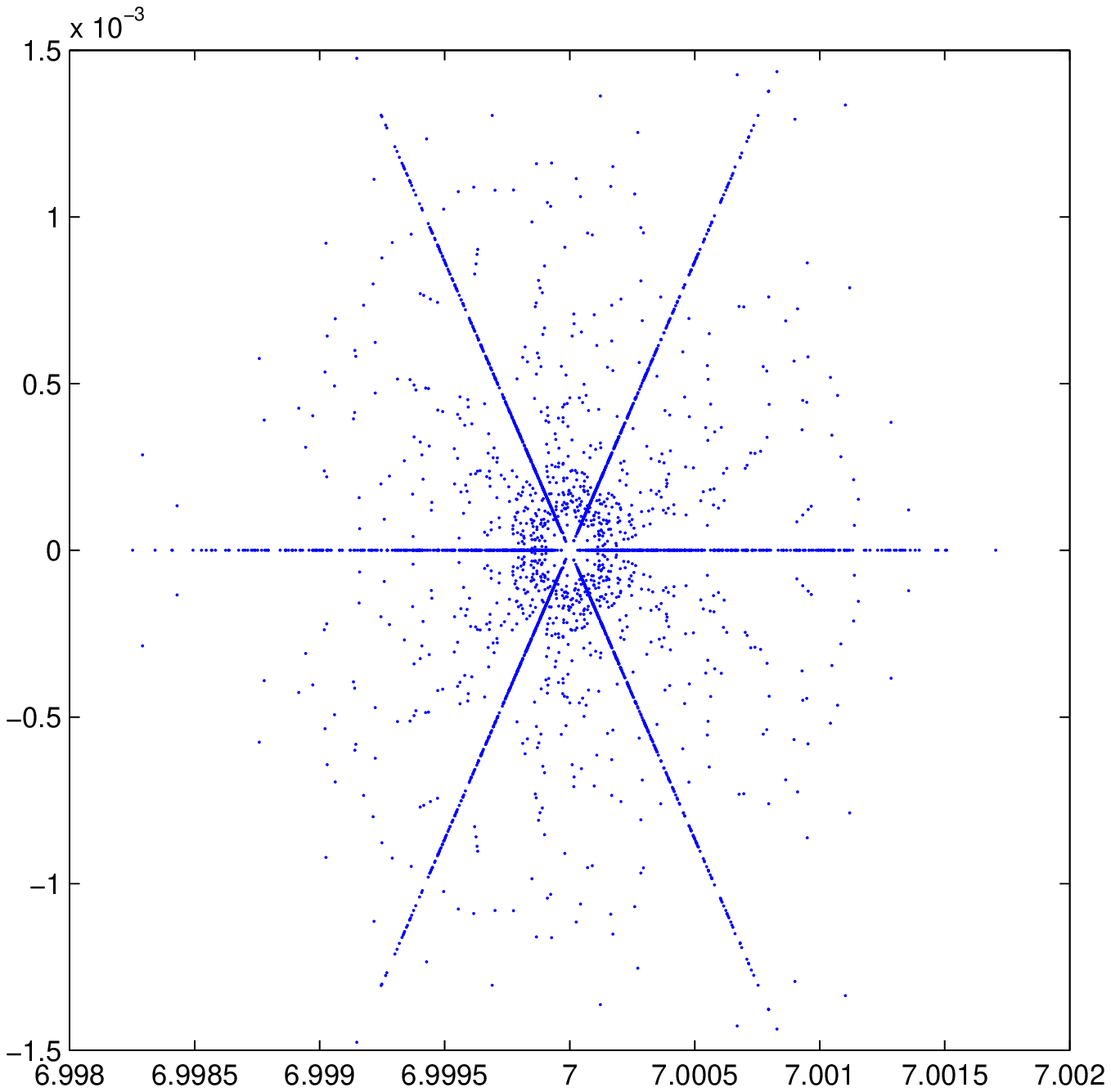} &
\includegraphics[width=0.25\textwidth]{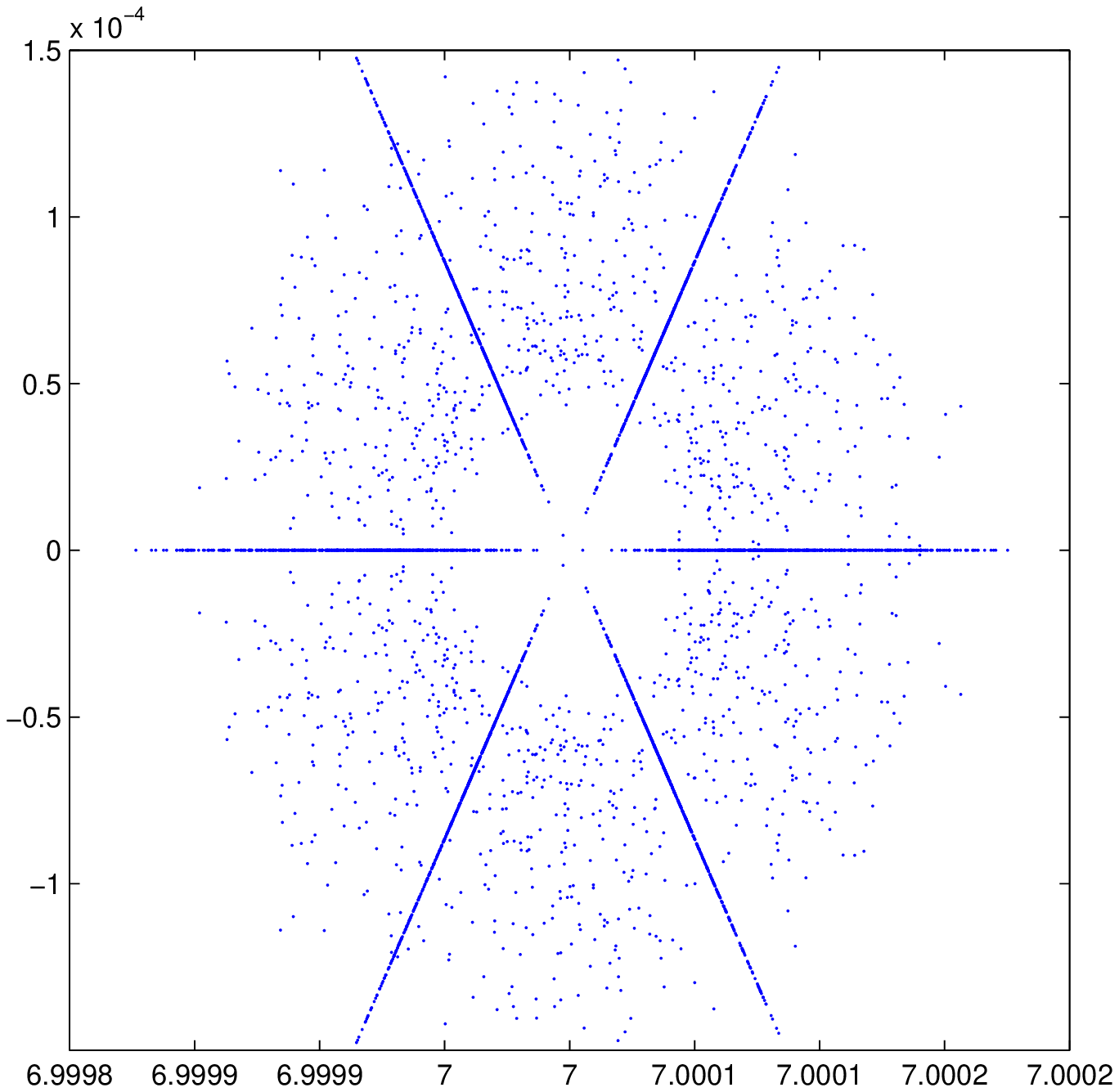} \\
\end{tabular}
\end{center}
\caption{Superimposed spectra for assessing the Jordan form
of the Brunet matrix. Two Jordan blocks of multiplicity 3 are
observed for eigenvalue 7, at different (left, right) perturbation levels.}
\label{fig:jordan}
\end{figure}

\subsubsection*{Qualitative assessment of Jordan forms}
A recent development has been the acceptance of a qualitative approach
to Jordan structure determination, proposed by Chaitin-Chatelin and
Frayss\'{e}~\cite{precise}. This approach does not employ the staircase
idea and, instead, exploits a semantics of eigenvalue perturbations to
infer multiplicity. This leads to a geometrically intuitive algorithm that
can be implemented using SAL.

Consider a matrix that has eigenvalues $\lambda_1, \lambda_2,
\cdots, \lambda_n$ with multiplicities $\rho_1, \rho_2, \cdots, \rho_n$
(resp). Any attempt at finding the eigenvalues (e.g., determining
the roots of the characteristic polynomial) is intrinsically subject
to the numerical analysis dogma: the problem being solved will
actually be a {\it perturbed} version of the original problem. This allows
the expression of the {\it computed} eigenvalues in terms of perturbations
on the actual eigenvalues. It can be easily seen that the computed
eigenvalue corresponding to any $\lambda_k$ will be distributed on
the complex plane as:
$$\lambda_k + |\Delta|^{1\over{\rho_k}} e^{{i\phi}\over{\rho_k}}$$
where the phase $\phi$ of the perturbation $\Delta$ ranges over \{$2\pi, 4\pi,
\ldots, 2\rho_k \pi$\} if $\Delta$ is positive and
over \{$3\pi, 5\pi, \ldots, 2(\rho_k+1) \pi$\} if $\Delta$ is negative. The
insight
in~\cite{precise} is to {\it superimpose} numerous such perturbed
calculations graphically so that the aggregate picture reveals the $\rho_k$ of
the eigenvalue $\lambda_k$. Notice that the phase variations
imply that the computed eigenvalues will be lying on the
vertices of a regular polygon centered on the {\it actual} eigenvalue
and where the number of sides is {\it two times} the multiplicity of the
considered eigenvalue (this takes into account both positive and
negative $\Delta$). Since the diameter of the polygon is influenced
by $\Delta$, iterating this process over many $\Delta$ will lead to a
`sticks' depiction of the Jordan form.

To illustrate, we choose a matrix whose computations will
be more prone to finite precision errors. Perturbations on
the 8-by-8 Brunet matrix~\cite{precise} with Jordan structure
$(-1)^1 (-2)^1 (7)^3 (7)^3$ induce the superimposed structures shown in
Fig.~\ref{fig:jordan}. The left part of Fig.~\ref{fig:jordan} depicts
normwise relative perturbations in the scale
of $[2^{-50},2^{-40}]$. The six sticks around the eigenvalue at 7
clearly reveal that its Jordan block is of size 3. The
other Jordan block, also centered at 7, is revealed if we conduct
our exploration at a finer perturbation level. Fig.~\ref{fig:jordan}
reveals the second Jordan block using perturbations in the
range $[2^{-53},2^{-50}]$. The noise in both pictures is a consequence
of (i) having two Jordan blocks with the same size, and (ii)
a `ring' phenomenon studied in~\cite{edelman-ma}; we do
not attempt to capture these effects in this paper.

\begin{figure}[t]
\begin{center}
\includegraphics[width=2in]{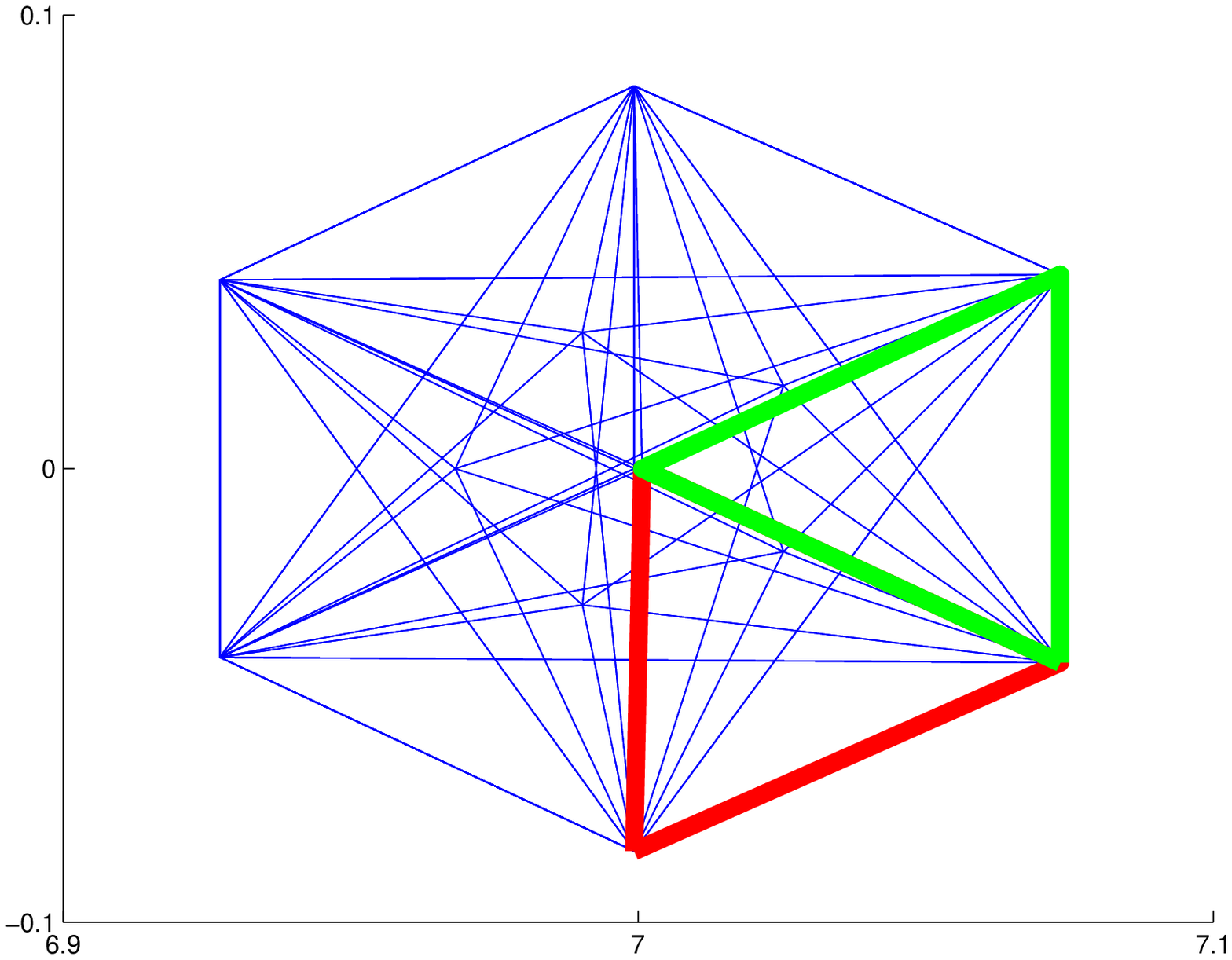}\hspace*{0.25in}\includegraphics[width=2in]{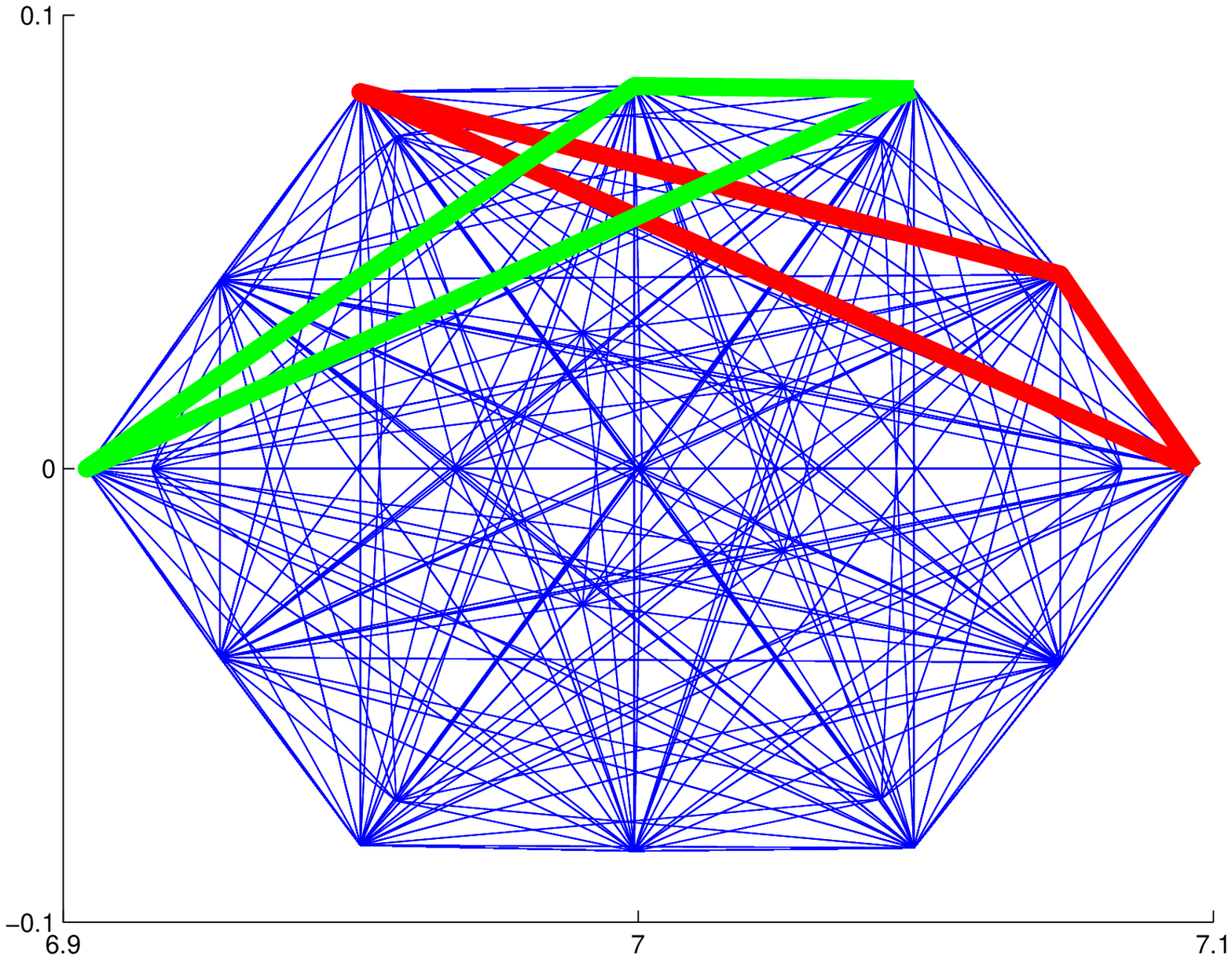} \\
\vspace*{0.1in}
\includegraphics[width=2in]{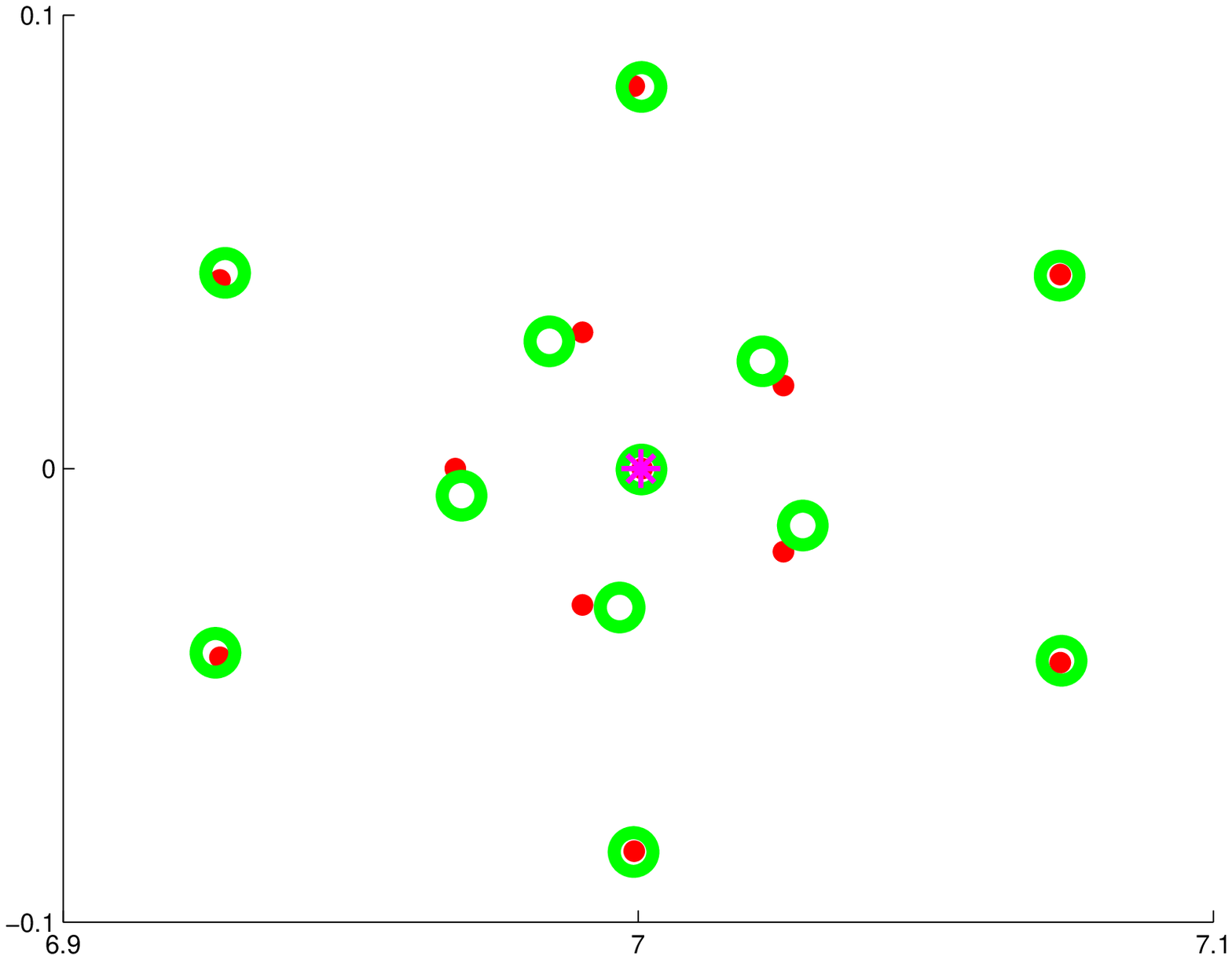}\hspace*{0.25in}\includegraphics[width=2in]{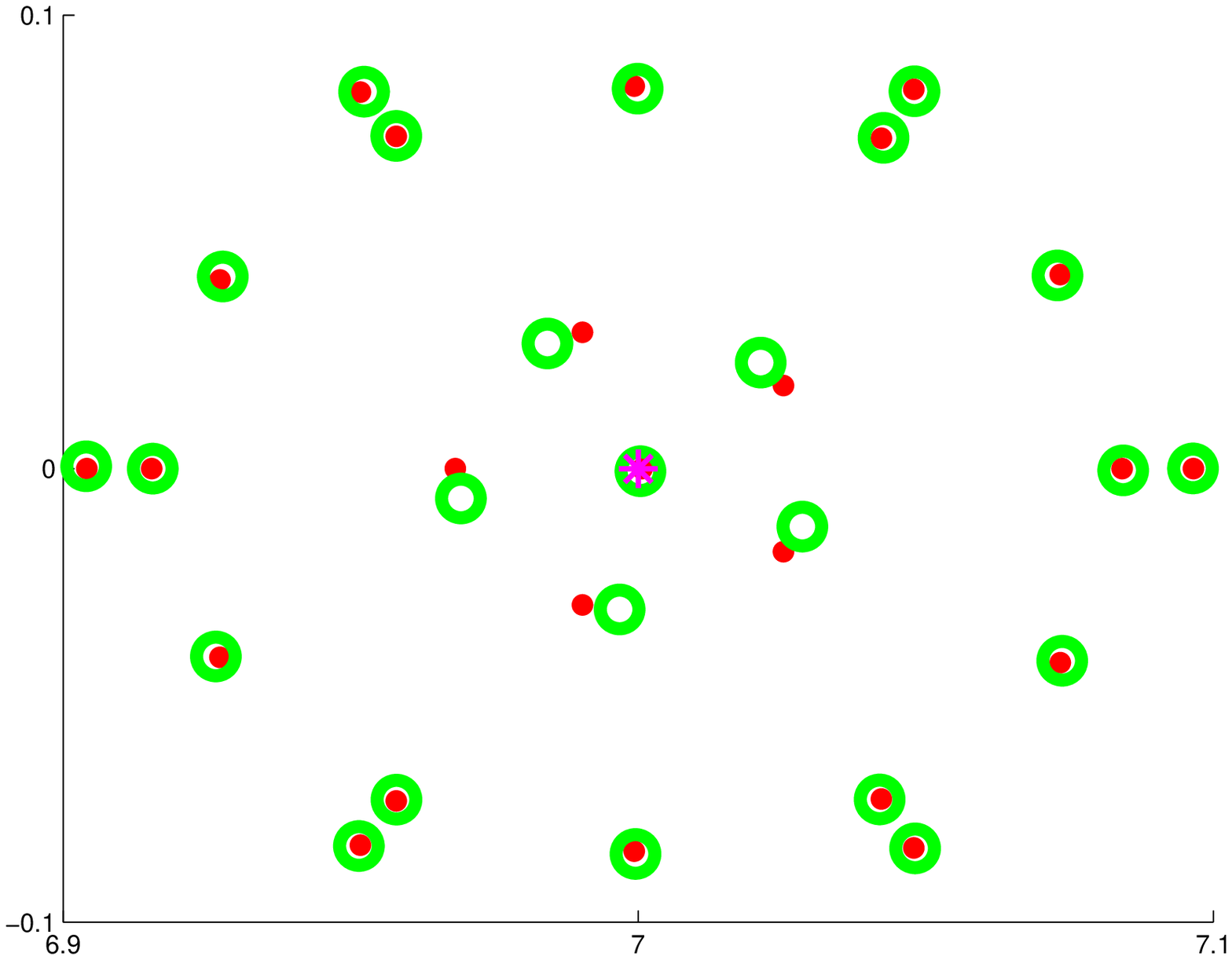} \\
\end{center}
\vspace*{-\baselineskip}
\caption{Mining Jordan forms from (left) a small sample set, and (right) 
large sample set. (top) Approximately congruent triangles. (bottom)
Evaluation
of correspondence of rotated triangles in terms of match 
%
between original (red dots) and rotated (green circles) samples.}
\label{fig:star-demo}
\end{figure}

\subsubsection*{Data Mining and Sampling Methodology}
For this study, we collect data by random normwise perturbations 
in a given region and a SAL program determines 
multiplicity by detecting symmetry correspondence in the samples.  The first
aggregation level collects the samples for a given perturbation into
triangles.  The second aggregation level finds congruent triangles via
geometric hashing~\cite{hash}, and uses congruence to establish
an analogy relation among triangle vertices.  This relation is then abstracted
into a rotation about a point (the eigenvalue), and evaluated for whether
each point rotates onto another and whether matches define regular
polygons. A third level then compares rotations across
different perturbations, re-visiting perturbations or choosing new
perturbations in order to disambiguate (see Fig.~\ref{fig:star-demo}). 
The end result of this
analysis is a confidence measure on models of possible Jordan forms.
Each model is defined by its estimate of $\lambda$ and $\rho$ (notice that
we are working only within one region at a time). The measure
$M$ was defined to be the joint probability distribution over the space of $\lambda$ and $\rho$.

\subsubsection*{Experimental Results}
Since our Jordan form computation treats multiple perturbations (irresp. of level)
as {\it independent} estimates of eigenstructure, the idea of sampling
here is not `where to collect,' but `how much to collect.' The goal of
data mining is hence to improve our confidence in model evaluation.
We organized data collection into rounds of 6-8 samples each,
varied a tolerance parameter for triangle
congruence from 0.1 to 0.5 (effectively increasing the number of
models posited), and determined the number of rounds needed to
determine the Jordan form. As test cases, we used the set of matrices
studied in~\cite{precise}.
On average, our focused sampling approach required 1 round of data collection
at a tolerance of 0.1 and up to 2.7 rounds at 0.5.  Even with a large number
of models posited, additional data quickly weeded out bad models. 
Fig.~\ref{fig:star-demo} demonstrates this mechanism on the Brunet matrix 
discussed above for two sets of sample points. 
To the best of our knowledge, this is the only known
known focused sampling methodology for this domain; we hence are unable to
present any comparisons. However, it is clear that by harnessing domain knowledge
about correspondences, we have arrived at an intelligent sampling methodology that
resembles what a human would obtain by visual inspection.

\section{Discussion}
The presented methodology for mining in data-scarce domains has several
intrinsic benefits. First, it is based on a uniform vocabulary of operators that
can be exploited for a rich diversity of applications. Second, it
demonstrates a novel factorization to the problem of mining when data is scarce,
namely, formulating an experiment design methodology to clarify, disambiguate,
and improve confidences in higher-level aggregates of data. 
This allows us to bridge qualitative and quantitative
information in a unified framework. SAL programs thus uncover bottom-up structures in data
systematically and use difficulties encountered in this process (ambiguities,
lack of correspondences) to guide top-down selection of additional
data samples. By using knowledge of physical properties explicitly, our
approach can provide more holistic and explainable results than off-the-shelf data
mining algorithms. Third, our methodology can co-exist with
more traditional approaches to problem solving (numerical analysis, optimization)
and is not meant to be a replacement or a contrasting approach. This is amply
demonstrated in each of the two applications above, where connections with various
traditional methodologies have been carefully established.

The methodology makes several intrinsic assumptions which we only briefly mention
here. All of our applications have been such that the cause, formation, and
effect of the relevant physical properties are well understood. This is precisely what
allows us to act decisively based on higher-level information from SAL aggregates,
through the measure $M$. It also assumes that the problems that will be
encountered by the mining algorithm are the same as the problems for which 
it was designed. This is an inheritance from Bayesian inductive inference and
leads to fundamental limitations on what can be done in such a setting. For instance, if
new data does not help clarify an ambiguity, 
does the fault lie with the model (SAL higher-level aggregate) or with the 
data? We can summarize this problem by saying that the approach requires strong {\it a priori} information about
what is possible and what is not. 

Nevertheless, by advocating targeted use of domain specific knowledge and
aiding qualitative model selection, our methodology is more efficient at determining
high level models from empirical data. Together, SAL and our information-theoretic measure $M$
encapsulate knowledge about physical properties and this is what makes our
methodology a viable one for data mining purposes.
In future we aim to characterize more formally
the particular forms of domain knowledge that help overcome sparsity and
noise in scientific datasets. 

It should be mentioned that while the two studies formulate their sampling objectives differently, they
are naturally supported by the SAL framework:
\begin{itemize}
\item (pockets) Where should I collect data in order to mine the pockets with high
confidence?
\item (Jordan forms) How much data should I collect in order to determine the right
Jordan form with high confidence?
\end{itemize} 
One could imagine extending our framework to also take into account the expense
of data samples. If the cost of data collection is non-uniform across the domain, then
including this in the design of our functional will allow us to tradeoff the cost
of gathering information with the expected improvement
in problem solving performance. This area of data mining is referred to as {\it active
learning.} 

Data mining can sometimes be a controversial term in a discipline that is used to
mathematical rigor; this is because it often used synonymously with `lack of a hypothesis
or theory.' We hope to have convinced the reader that this need not be the case and
that data mining can indeed be sensitive to knowledge about the domain, especially
physical properties of the kind we have harnessed here. As data mining
applications become more prevalent in science, the need to incorporate {\it a priori}
domain knowledge will only become more important.

\end{document}